\documentclass[fleqn,10pt]{wlscirep}
\usepackage{amssymb}
\usepackage{amsfonts}
\usepackage{amsmath}
\usepackage{amssymb}
\usepackage{amstext}
\usepackage{bm}
\usepackage{mathtools}
\usepackage{graphicx}
\usepackage{hyperref}
\usepackage{lineno}
\newcommand{\subheader}[1]{\noindent\textbf{#1}\newline\noindent}
\graphicspath{ {./figures/} }

\newcommand{\ent}[1]{e\scalebox{0.55}[0.55]{\ifthenelse{\equal{#1}{+}}{\raisebox{0.65ex}[0pt]{$\mathbf{+}$}}{\raisebox{0.65ex}[0pt]{$\mathbf{-}$}}}}

\newcommand{\EX}{\mathbb{E}} 
\newcommand{\EXP}[2][x]{\EX_{#1}\left\{#2\right\}} 

\newcommand{\SX}{\mathrm{Std}} 
\newcommand{\SXP}[2][x]{\SX_{#1}\left\{#2\right\}} 

\newcommand{\LX}{\mathcal{L}} 
\newcommand{\NX}[1]{\left\lVert #1 \right\rVert}

\newcommand{\FT}{\mathcal{F}} 

\newcommand{\drm}{\mathrm{d}}

\newcommand{\dr}{\drm r} 

\newcommand*\circnum[1]{\raisebox{.5pt}{\textcircled{\raisebox{-.9pt} {#1}}}}

\title{Deep convolutional neural networks to restore single-shot electron microscopy images}

\author[1,2,*]{I. Lobato}
\author[1,2]{T. Friedrich}
\author[1,2,*]{S. Van Aert}
\affil[1]{EMAT, University of Antwerp, Department of Physics, Groenenborgerlaan 171, B-2020 Antwerp, Belgium}
\affil[1]{NANOlab Center of Excellence, University of Antwerp, Department of Physics, Groenenborgerlaan 171, B-2020 Antwerp, Belgium}
\affil[*]{Ivan.Lobato@uantwerpen.be}
\affil[*]{Sandra.VanAert@uantwerpen.be}


\begin{abstract}
State-of-the-art electron microscopes such as scanning electron microscopes (SEM), scanning transmission electron microscopes (STEM) and transmission electron microscopes (TEM) have become increasingly sophisticated. However, the quality of experimental images is often hampered by stochastic and deterministic distortions arising from the instrument or its environment. These distortions can arise during any stage of the imaging process, including image acquisition, transmission, or visualization. In this paper, we will discuss the main sources of distortion in TEM and S(T)EM images, develop models to describe them and propose a method to correct these distortions using a convolutional neural network. We demonstrate the effectiveness of our approach on a variety of experimental images and show that it can significantly improve the signal-to-noise ratio resulting in an increase in the amount of quantitative structural information that can be extracted from the image. Overall, our findings provide a powerful framework for improving the quality of electron microscopy images and advancing the field of structural analysis and quantification in materials science and biology.
\end{abstract}

\begin{document}

\flushbottom
\maketitle
\thispagestyle{empty}

\section*{INTRODUCTION}\label{sec:intro}
The quality of modern electron microscopes, such as scanning electron microscopes (SEM), scanning transmission electron microscopes (STEM), and transmission electron microscopes (TEM), has greatly improved. However, the quality of the experimental images produced by these instruments is often compromised by stochastic and deterministic distortions arising from the instrument or its environment \cite{NCDKEA04,Joy2005,KFBEA08}. These distortions can occur during the acquisition, transmission, or reproduction of the image. Despite technical improvements in the design of high-performance electron microscopes \cite{NCDKEA04,Joy2005,KFBEA08,MKTGFW06}, the presence of these distortions in the recorded images may hinder the extraction of quantitative information from the samples under study \cite{Jones2017}.

In TEM, images are acquired in a single shot using parallel acquisition. Here, the main sources of distortions are the detector noise, which is a combination of counting noise associated with the uncertainty of photon/electron detection, dark current noise resulting from statistical variation in the number of thermally generated electrons within the detector, and readout noise resulting from the electronics that amplifies and digitizes the charge signal. Other sources of distortions for TEM include X-ray noise, which is produced by X-rays that saturate one or more nearby pixels as they pass through the detector \cite{Zuo2000,Faruqi2000}, and dead pixel noise, which is caused by permanently damaged pixels on the sensor and often appears as black spots in the recorded images.

In S(T)EM, images are formed pixel by pixel by scanning a convergent electron beam across the sample and detecting the scattered, back-scattered or secondary electrons at each point. The main sources of distortions are the detector noise, which is a combination of shot noise hitting the scintillator, Gaussian noise resulting from the photomultiplier tube (PMT) \cite{Ishikawa2014}, and readout noise from the electronics that amplifies and digitizes the electron signals. Unlike TEM imaging, the serial nature of SEM and STEM can introduce additional distortions into the resulting images due to time delays between measurements. At high doses, the main source of nonlinear distortion is the probe's fly-back time, where data collection pauses until scanning on the next line resumes. This produces a net two-dimensional random displacement of the pixel row known as horizontal and vertical scan distortion. These nonlinear distortions can often be corrected using iterative algorithms that require a series of images \cite{BBBDSV14, JYPMABCN15} or a single image with a high-resolution periodic structure \cite{LJPN13, BBLR12}. Moreover, S(T)EM images obtained through high-speed scans (dwell time $<1\mu s$ \cite{BRGBS10}) may display a non-uniform scan speed along individual scan lines resulting in a smearing effect that produces another type of nonlinear distortion. While these distortions can be partly compensated for periodic structures \cite{BRGBS10}, they cannot be fully compensated for arbitrary specimens. Other types of distortion include row-line noise, which is caused by the detector's non-response over a few pixels, and X-ray noise, which is produced by X-rays hitting the detector. These distortions can reduce the signal-to-noise ratio (SNR) and limit the amount of retrievable information about the electron-specimen interaction. Moreover, they can cause translation, shear, rotation, expansion, or contraction of the entire image. Although increasing the beam current or acquisition time can improve the SNR, it can also increase other types of distortion, such as shear or rotation. Moreover, it is unsuitable for beam-sensitive materials or for dynamic imaging requiring a short exposure time for each frame. Lowering the electron dose can also decrease the quality of the recorded images and limit the reliability of structural information extracted from them.

Various algorithms have been developed to improve the SNR of electron microscopy (EM) images, including spatial filters such as median filters, Gaussian filters, Bragg filters, and Wiener filters \cite{JCVD89, LDM96, KTS12}. More complex methods for denoising EM images include non-linear iterative Wiener filtering algorithms \cite{HD15} and block matching \cite{Dabov2007, MBDVYB15} although they can be computationally intensive. Another option for improving the SNR is to average a series of registered frames, using either rigid \cite{KAYNMI10} or non-rigid \cite{BBBDSV14, JYPMABCN15} registration methods. However, these methods require a high overall electron dose and repeated recordings of the material. In addition, EM images often exhibit a combination of different types of distortions due to several factors including the instrument environment, scan instabilities, scan speed, and dose. Therefore, there is a need for image restoration algorithms specifically designed for single-shot EM images.

In recent years, machine learning methods based on artificial neural networks, particularly convolutional neural networks (CNNs), have become the state-of-the-art approach for various tasks such as image classification \cite{Huang2017}, image segmentation \cite{Chen2018}, image denoising \cite{Kim2019}, image restoration \cite{Cheng2018}, image deconvolution \cite{Luo}, and image super-resolution \cite{Wang2019}. These methods, which involve adjusting the weight connections between neurons during training, have been made possible by the development of techniques such as the Rectified Linear Unit (ReLU) \cite{VNEH10}, dropout regularization \cite{SHKSS14}, batch normalization \cite{Ioffe2015}, and improvements in GPU technology. While CNN-based approaches have achieved strong performance in denoising specific types of EM images, they are limited by their reliance on small simulated or experimental datasets and incomplete modelling of the various types of noise present in experimental EM data \cite{Wang2020a, Wang2020b, Mohan2022, Lin2021, Kaiser2023}. To the best of our knowledge, there is currently no algorithm that can effectively compensate for all types of distortion in a single-shot EM image without requiring retraining and regardless of the sample being studied.

In this study, we use a machine learning approach to restore EM images using a Concatenated Grouped Residual Dense Network (CGRDN) and a combination of loss functions and a generative adversarial network (GAN) \cite{Mirza2014}. This approach not only learns an end-to-end mapping between distorted and undistorted EM images, but also a loss function to train this mapping. Since we only have access to distorted data experimentally, we generate undistorted and distorted EM images by applying all distortions that can be corrected on single-shot EM images. By training the neural network to produce an undistorted output regardless of the level and combination of distortions in the input, it implicitly learns to detect and repair the distortions. This approach demonstrates impressive results for restoring both periodic and non-periodic specimens with different combinations of severe distortions. Importantly, the results show that both peak positions and intensities in atomic resolution images can be reliably determined. In addition, the restoration time is only of the order of seconds for a 2kx2k image.

\section*{RESULTS AND DISCUSSION}\label{sec:results}
Electron microscopy techniques, namely SEM, STEM, and TEM, exhibit distinct sources of noise and variations in their features at both low and high resolution. Hence, we have trained our network architecture on six diverse datasets comprising low-resolution (LR) and high-resolution (HR) images for each microscopy modality. Our findings indicate that the best performance is achieved by training separate networks for LR and HR features, particularly at low doses, where the network can utilize the specific feature distribution acquired during the training phase. Detailed implementation and training information is provided in the supplementary material. Our study mainly focuses on HR-STEM, a widely used technique for the analysis and quantification of atomic structures.

\subsection*{Ablation study and comparison to state-of-the-art  algorithms}\label{sec:Ablation_study}
To improve the performance of a neural network, it is important to choose the right values for the hyperparameters. These values can affect the network's ability to minimize errors, run quickly, and fit within certain hardware constraints. In our case, we want the network to be able to process images of size $1024 \times 1024$ in less than one second, and we want to be able to run it on an Nvidia Volta GPU card with 12GB of memory. To find the best hyperparameters for our needs, we perform an ablation study. This involves varying the network architecture and some of its hyperparameters and measuring their effect on the $\LX_1$ error (see "Loss function" section). Since our hardware constraints limit the maximum number of residual dense blocks (RDB), grouped residual dense blocks (GRDB), and batch size to 4, we will keep these values constant at their maximum value. All other parameters of our generator are defined in the "Network architecture" section and will be kept constant unless otherwise specified. A grid search is used to find the optimal values for the learning rate and loss weighting parameters.

\begin{figure}[!ht]
	\centering \includegraphics[width=8cm]{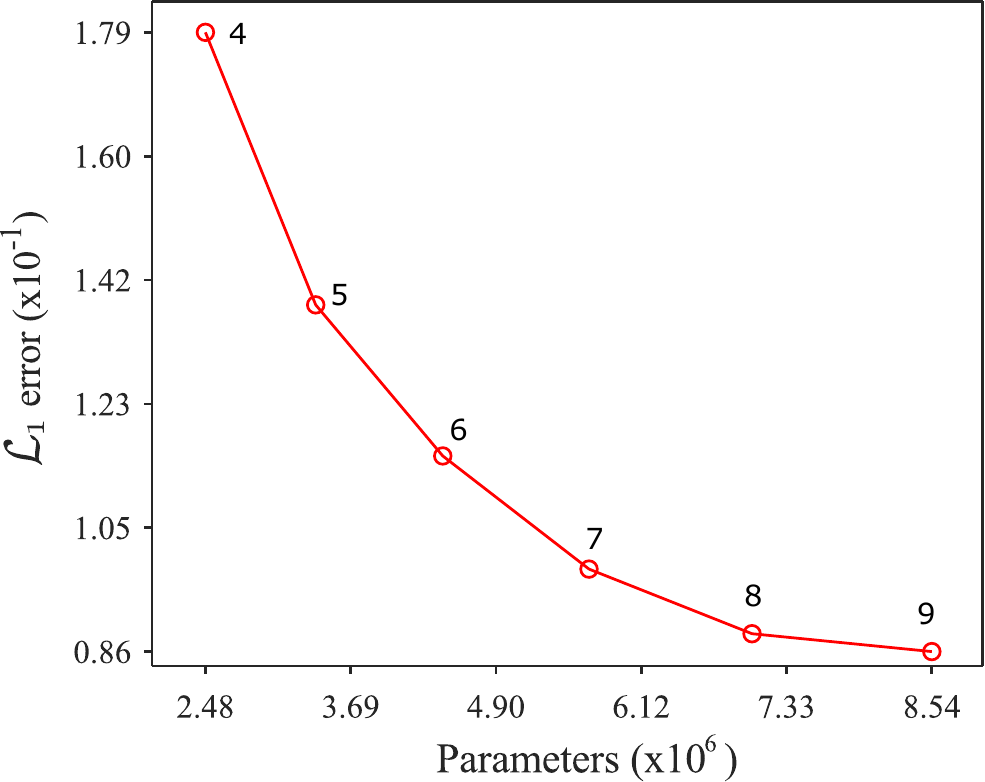}
	\caption{Ablation study of the CGRDN architecture based on $\LX_{1}$ metric as a function of the size of the model. The number of layers $n_{lay}$ is indicated next to each data point.}
	\label{fig:fig_ablation_study}
\end{figure}

In the first part of this ablation study, we focus on the performance of the network when the number of convolutional layers $n_{lay}$ within the RDB increases. Figure \ref{fig:fig_ablation_study} shows the reduction of the $\LX_1$ error when the number of layers and network parameters increases. This is expected since a deeper network can improve the performance of the model by increasing the number of parameters and allowing the model to learn more complex features.
We would like to highlight that our hardware constraints only allow us to use a maximum of 9 layers for $n_{lay}$. Nonetheless, we observed that the $\LX_1$ error starts to reach a plateau for $n_{lay}=9$, indicating that increasing the number of layers may not lead to substantial performance improvements.

Furthermore, we compared the performance of three different image denoising architectures: the Grouped Residual Dense Network (GRDN) \cite{Kim2019}, the Multi-resolution U-Net (MR-UNET) \cite{Wang2020b}, and our proposed architecture, CGRDN. We assessed the performance of these architectures using the well-known peak signal-to-noise ratio (PSNR), which is defined as:

\begin{equation}
    PSNR = 10\log_{10}\left(\frac{MAX^2}{MSE}\right),
\end{equation}
where $MAX$ denotes the maximum possible pixel value of the images, and $MSE$ represents the mean squared error between the distorted and undistorted images. However, it is important to note that PSNR only measures the pixel-wise differences between the original and reconstructed images and does not account for other crucial factors such as visual perception and structural similarity. The GRDN architecture was previously ranked first in terms of PSNR and structure similarity index in the NTIRE2019 Image Denoising Challenge. The MR-UNET extends the functionality of the decoder in a U-Net \cite{Ronneberger2015} by adding additional convolutional layers to the hidden layers in order to produce coarse outputs that match low-frequency components. The results of our comparison are summarized in Table \ref{tab:table1}, which shows the number of parameters and the resulting PSNR for each architecture and show that the GRDN and CGRDN are more efficient architectures because they require approximately 7 times fewer parameters than the MR-UNET, while still achieving a higher PSNR. It is interesting to note that our CGRDN architecture achieved a higher PSNR than the GRDN, while only requiring an additional 20,000 parameters.
\begin{table}[h!]
	\begin{center}
		\caption{PSNR denoising performance comparison of different network architectures.}
		\begin{tabular}{lll} 
			\hline
			\text{Method} & \text{$\#$ parameters} & PSNR\\
			\hline
			\text{MR-UNET} \cite{Wang2020b} & 51.7M & 36.70dB\\
			\text{GRDN} \cite{Kim2019} & 7.02M & 36.90dB\\
			\text{CGRDN} \text{this work} & 7.04M & 36.96dB\\	
			\hline
		\end{tabular}
		\label{tab:table1}
	\end{center}
\end{table}

\begin{figure}[!ht]
	\centering \includegraphics[width=\linewidth]{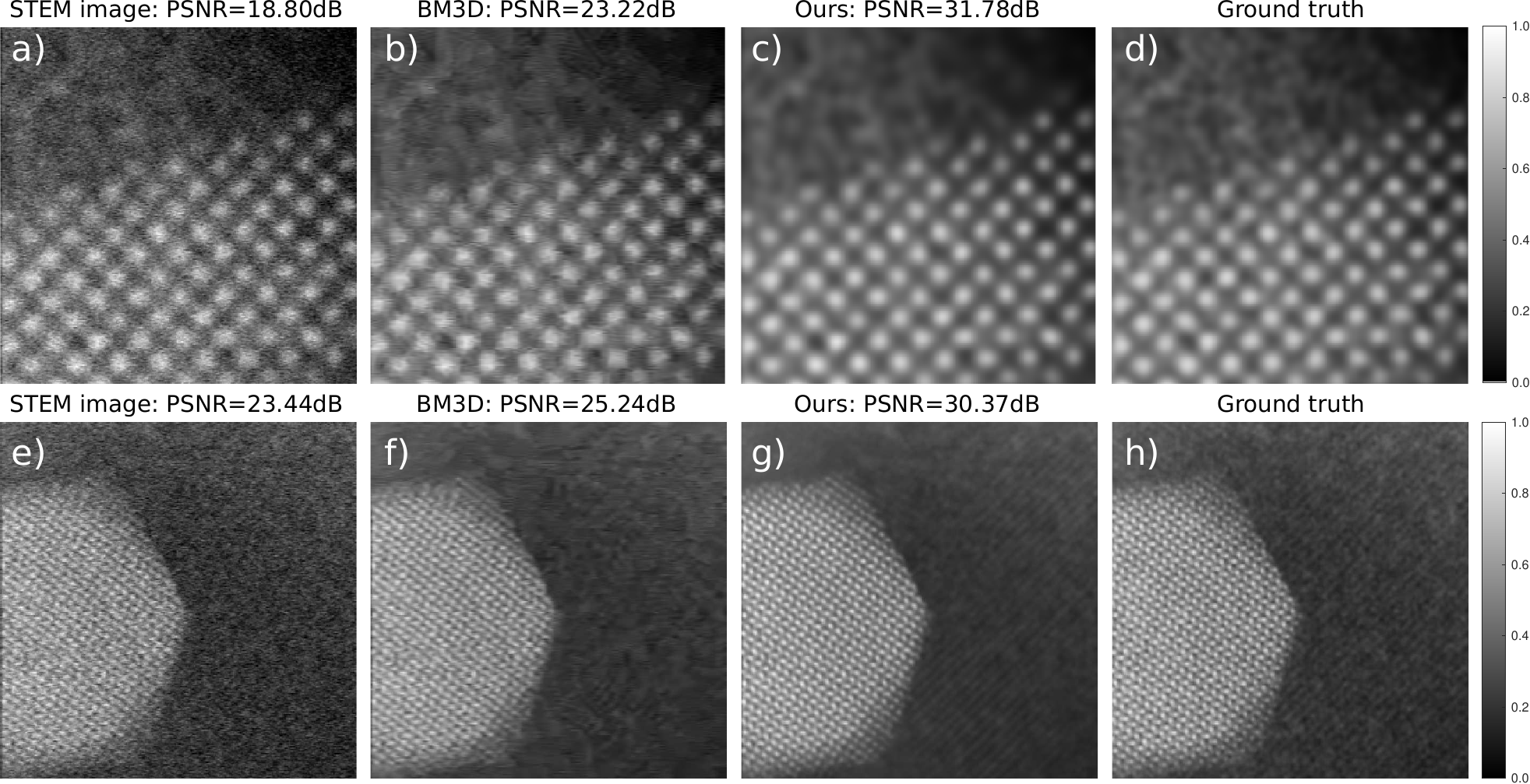}
	\caption{CNN restoration results compared with BM3D in terms of PSNR for two random simulated STEM specimens using standard experimental noise values.}
	\label{fig:bm3d}
\end{figure}

We also compared the performance of our image restoration network to the Block-matching and 3D filtering (BM3D) \cite{Dabov2007} algorithm in terms of PSNR. BM3D is a widely used technique for removing noise from images through a process called denoising. It segments the image into overlapping blocks and identifies similar patterns among them to estimate the original image and reduce noise. BM3D has demonstrated effectiveness in denoising images with high levels of noise and serves as a benchmark for image denoising algorithms in image processing. The average PSNR of BM3D and our network on the validation dataset was $30.45$ dB and $36.96$ dB, respectively. These results demonstrate that our network outperforms BM3D by a significant margin of $6.51$ dB. Figure \ref{fig:bm3d} illustrates the performance of our network and BM3D on two randomly generated, high-resolution STEM images with standard experimental noise values. These images were simulated using the procedure outlined in the "Data generation" section. The figure displays the original distorted images (a)\&(e) and undistorted images (d)\&(h), as well as the denoised output from BM3D (b)\&(f) and the restored output from our network (c)\&(g).
These results demonstrate that our image restoration network significantly enhances image quality, as measured by PSNR. However, it is noteworthy that PSNR is not always a reliable indicator of image quality since it merely measures pixel-wise differences between original and reconstructed images and overlooks other critical factors such as visual perception and structural similarity. Hence, it is crucial to employ various image quality metrics, along with PSNR, to obtain a more comprehensive evaluation of the performance of image restoration techniques.

\subsection*{Atomic structure quantification}\label{sec:results_quant}
\begin{figure}[!ht]
	\centering \includegraphics[width=\linewidth]{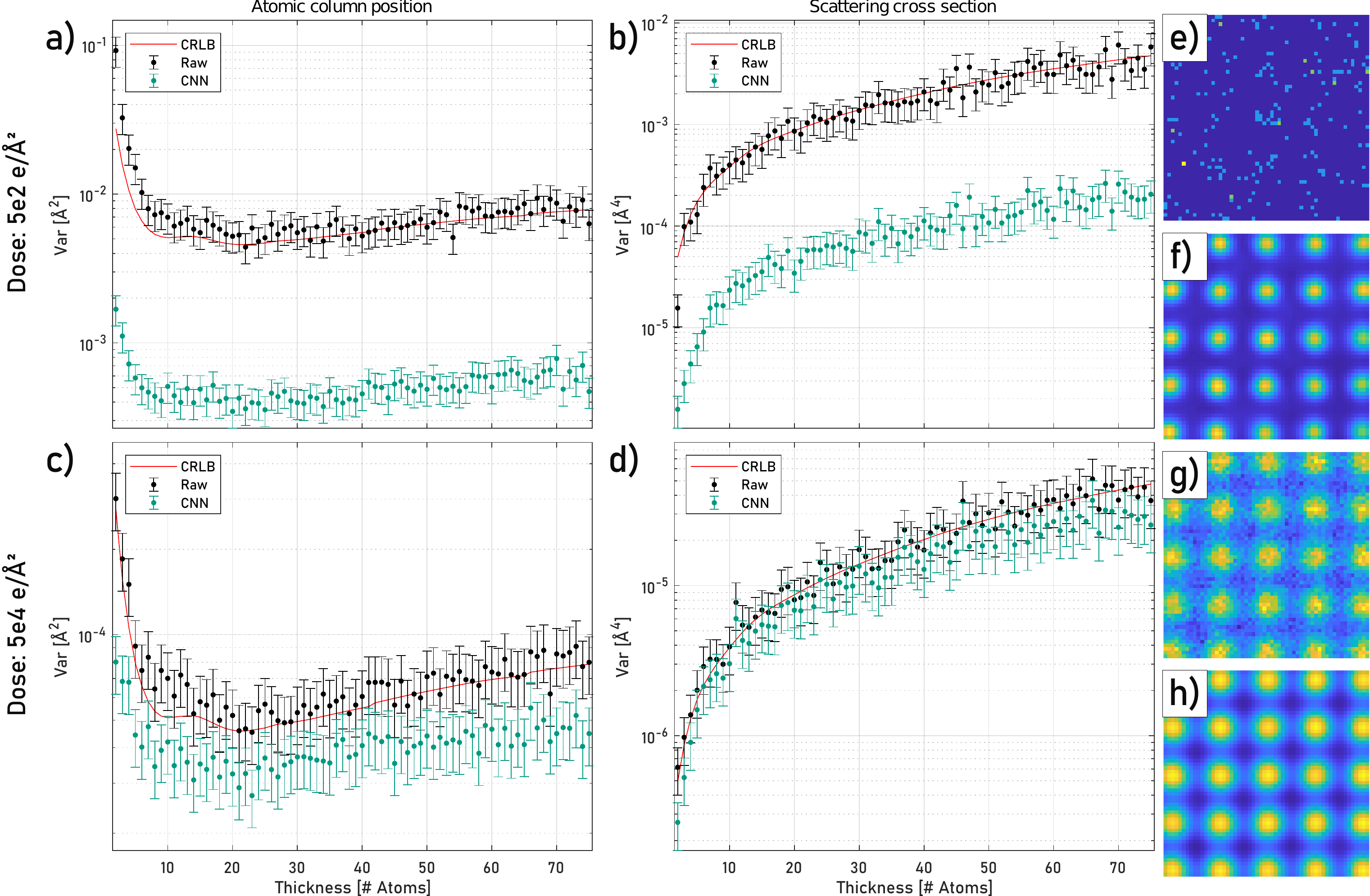}
 \caption{Precision of atomic column position and SCS-measurements over a series of Pt-bulk samples with a thickness varying from 2-75 atoms together with their 95\% confidence intervals. (a) Precision of the atomic column locations for a dose of 5e2 $e/\text{\AA}^2$. (b) Precision of SCS measurements for a dose of 5e2 $e/\text{\AA}^2$. (c) Precision of atomic column locations for a dose of 5e4 $e/\text{\AA}^2$. (d) Precision of SCS measurements for a dose of 5e4 $e/\text{\AA}^2$. (e) Example of a raw STEM image at z=2 and dose=5e2 $e/\text{\AA}^2$. (f) Example of a restored STEM image at z=2 and dose=5e2 $e/\text{\AA}^2$. (g) Example of a raw STEM image at z=75 and dose=5e4 $e/\text{\AA}^2$. (h) Example of a restored STEM image at z=75 and dose=5e4$e/\text{\AA}^2$.}
 \label{fig:crlb}
\end{figure}

While the CNN was trained to restore images of a wide variety of imaging modes, STEM is of particular interest since it is routinely used for the quantification of atomic structures \cite{VanAert2009, Martinez2014,DeBacker2016} in terms of atomic column positions and their corresponding scattering cross sections (SCS), which allows us to study the impact of the proposed image restoration method quantitatively. The probe position integrated scattering cross section, short SCS, in atomic resolution STEM images is defined as the integrated intensity of an atomic column, which is typically modelled as a 2D gaussian function. Since the SCS scales with the atomic number $\approx Z^{1.7}$\cite{Krivanek2010a, Yamashita2018a} and mostly increases monotonically with thickness for large collection angles, it is routinely used for atom counting. The evaluation of the effect of image restoration on the quantitative assessments of STEM images is done in three complementary approaches, using MULTEM \cite{Lobato2015, Lobato2016} to create multislice simulations and the StatSTEM software for all model fittings \cite{DeBacker2016}. All evaluations are based on 100 distortion/noise realisations for each dose setting.
\begin{enumerate}
    \item We demonstrate the effect of image denoising with an idealised setup in analogy to the study conducted in reference \cite{DeBacker2016}, where the precision of the determination of the location and SCS of an atomic column was determined over a wide range of signal-to-noise-ratios (SNRs) using pure Poisson noise. This setting allows the comparison to the theoretical limits of variance for unbiased estimators, the so-called Cramér–Rao-Lower Bounds(CRLBs). The simulated STEM dataset is a bulk Pt crystal in [001] orientation and contains STEM images over 75 depth sections with unit cell spacing in z-direction.
    \item A more practical example, that includes crystal irregularities, is chosen to determine the impact of a combination of noise, scan-line-distortions and fast-scan distortion. In this case, we evaluate the mean absolute error (MAE) for atomic column positions and the mean absolute percentage error (MPE) for the SCSs of atomic columns, as well as the variance of these measurements. This serves to show in particular the independence of the approach on the structural periodicity for atomic-resolution STEM images.
    \item For a simulated Pt-nanoparticle it is demonstrated that distortion correction yields not only a more accurate localisation of atomic columns but also enables more reliable atom counting.
\end{enumerate}
The simulation settings for all samples are tabulated in the supplementary information. The results of the first study are shown in figure \ref{fig:crlb}. Examples of the underlying STEM images are given for the extremes of SNRs (i.e. smallest thickness and lowest dose and largest thickness and highest dose) for raw and restored images in panels (e), (f), (g) and (h). Comparing figure \ref{fig:crlb}(e) and (f) it can be seen visually that even at a very low dose, the CNN can recover the underlying structure faithfully. This effect is measurable both in terms of the precision with which atomic columns can be located, as well as in SCS measurement precision, and is particularly pronounced in the low dose range as illustrated in figure \ref{fig:crlb}(a) and (b). As the dose increases the precision of the structural measurements of raw and restored data converge eventually (figure \ref{fig:crlb}(c-d)). An interesting observation is that the theoretical precision limit given by the CRLB, can be overcome employing image restoration. This makes a strong point for using image restoration for quantitative studies, like atom counting or strain measurements in general.

\begin{figure}[!ht]
	\centering \includegraphics[width=\linewidth]{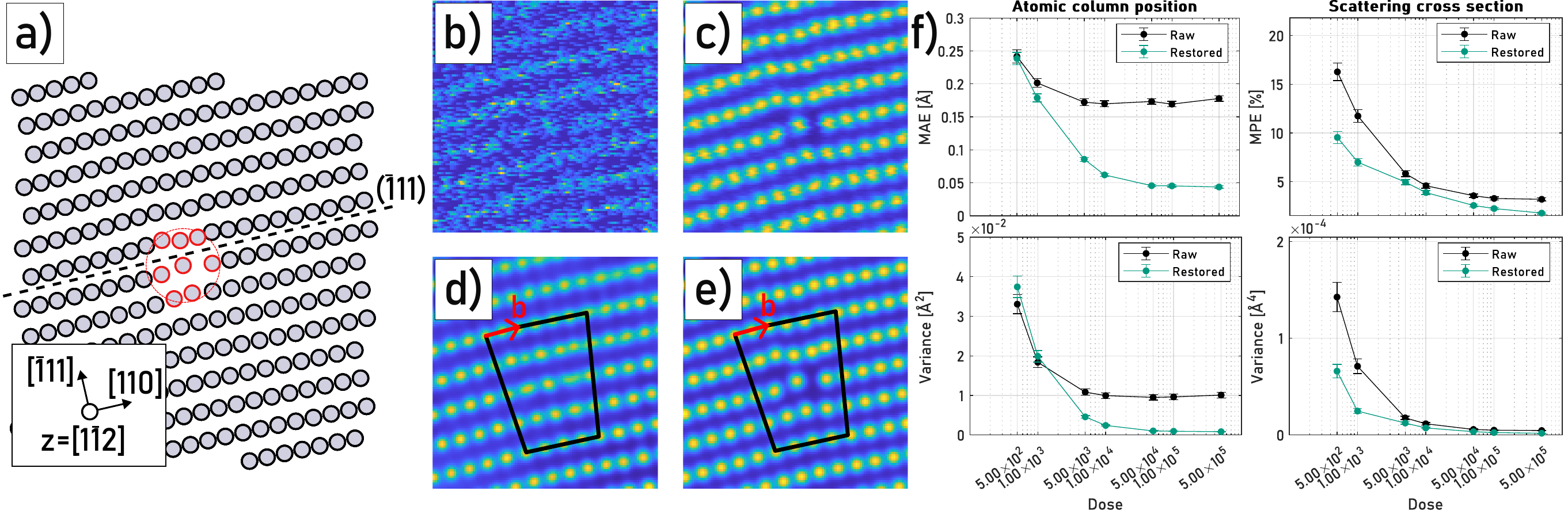}
    \caption{(a) Schematic of the Pt structure in [112] zone axis with a unit edge dislocation of Burgers vector $b=1/2[110]$ in the $(111)$ glide plane. (b) Corrupted raw HAADF STEM image with a dose of $5e2 e/\text{\AA}^2$. (c) Corrupted raw image with a dose of $5e5 e/\text{\AA}^2$.  (d) Restored image with a dose of $5e2 e/\text{\AA}^2$. (e) Restored image with a dose of $5e5 e/\text{\AA}^2$. (f) Quantification results for the atomic column positions and scattering cross sections of the atomic columns around the center of the edge dislocation (marked with red circles in panel (a)).}
    \label{fig:edge_dis}
\end{figure}

The restoration results in the first example arguably benefit from the underlying perfect crystal symmetry, which is why we test the CNN also for imperfect structures. The Pt-bulk model depicted in figure \ref{fig:edge_dis}(a) is in $[112]$ zone axis orientation, six unit cells thick and contains a unit edge dislocation of Burgers vector $b=1/2[110]$ in the $(111)$ glide plane; a dislocation commonly observed in fcc metals \cite{Hull2011}. The structure was created using the Atomsk software, which determines atom positions corresponding to the displacement fields predicted by the elastic theory of dislocations \cite{Hirel2015}. The simulated HAADF STEM images were subjected to varying noise levels from 5e2 $e/\text{\AA}^2$ to 5e4 $e/\text{\AA}^2$, and further corrupted by scan-line distortions as outlined in the "S(T)EM noise model" section. Example reconstructions for raw images at doses of 5e2~$e/\text{\AA}^2$ and 5e4~$e/\text{\AA}^2$ (figure \ref{fig:edge_dis}(b) and (c)) are shown in figure \ref{fig:edge_dis}(d) and (e), respectively. In the low-dose raw image individual atomic columns are hardly recognisable. Without the prior knowledge of the atomic column positions, any attempt of model fitting would have to overcome the challenge of performing reliable peak finding first, which is a factor not considered here. The reconstruction of this image (figure \ref{fig:edge_dis}(d)) on the other hand shows very clear peaks. A burgers circuit is superimposed on the image to highlight that despite the poor separation of columns in the raw image, the dislocation with its correct burgers vector $b$ is maintained, which means that the structure as a whole is retrieved correctly, albeit the individual column positions may not be fully accurate as can be seen in the mean absolute position error of the columns around the center of the dislocation (columns within the red circle in figure \ref{fig:edge_dis}(a)) for low doses shown in figure \ref{fig:edge_dis}(f). However, the error drops rapidly with increasing dose and shows a clear improvement against raw images. The position accuracy is therefore not only a result of denoising but also the result of the accurate correction of scan-line and fast-scan distortions. The comparatively high accuracy for the raw image fitting at low doses can be attributed to the fact that correct initial column positions are given for the fitting procedure. Since the column can hardly be located in the noisy images, the fitting algorithm on average does not move the position much away from this initial position. The CNN on the other hand reconstructs a clearly visible atomic column, but the available information in the underlying image is insufficient for accurate positioning. However, the proper retrieval of the dislocated atomic column at higher doses shows that the CNN is not by default just picking up on periodicity, but faithfully recovers the atomic structure also in the presence of non-periodic features in atomic resolution STEM images. 

\begin{figure}[!ht]
	\centering \includegraphics[width=\linewidth]{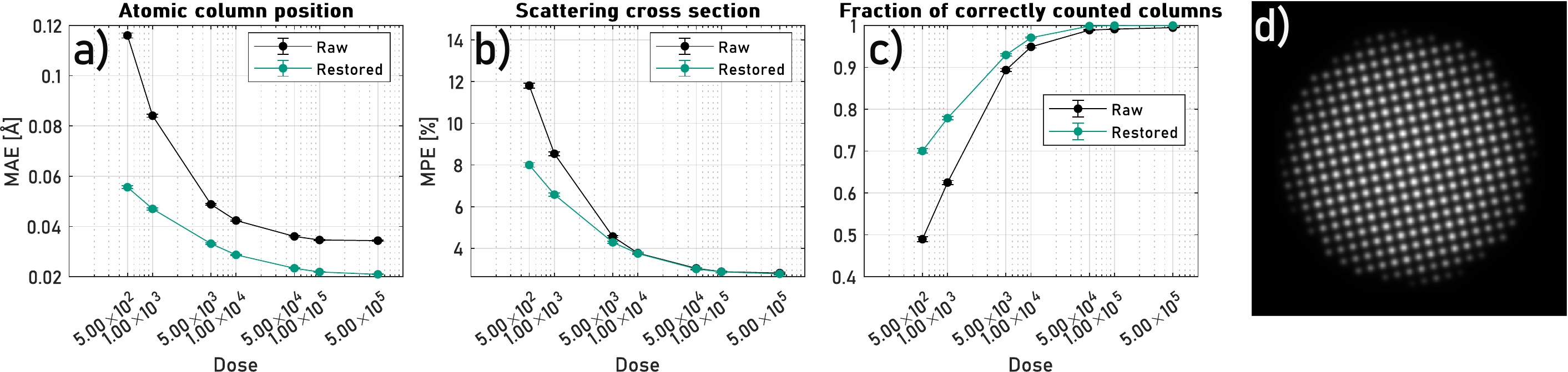}
    \caption{Quantification results for a spherical Pt nanoparticle with a diameter of 11 unit cells in [100] orientation. The values are based on all 333 atomic columns for 100 noise realisations. (a) The mean absolute error of the estimated atomic column positions. (b) The mean absolute percentage error of the fitted scattering cross sections, which are being used to estimate atom counts in each column. (c) The fraction of atomic columns with correctly estimated atom counts. }
    \label{fig:Pt_NP_results}
\end{figure}

Also the SCS measurements improve in accuracy by the restoration, which would translate directly into improvements for atom counting studies. An example of such an atom counting scenario is presented in figure \ref{fig:Pt_NP_results}. These results were obtained from a simulated spherical Pt nanoparticle with 11 unit cells in diameter in [100] zone axis orientation under the same distortion and noise parameters as given in the previous example. Atom counts were obtained by matching retrieved SCS values against simulated library values\cite{VanAert2013}. The improvement in column position measurements over all dose settings again indicates the proper correction of scan-line and fast-scan distortions. The improvement of SCS measurement accuracies, especially at low-dose conditions greatly decreases the chance of miscounting atoms in the structure, which in turn may be very beneficial e.g. for the reconstruction of 3D information from atom-counts \cite{ArslanIrmak2021, DeBacker2022}.

\subsection*{Experimental image restorations}
\begin{figure}[!ht]
	\centering \includegraphics[width=\linewidth]{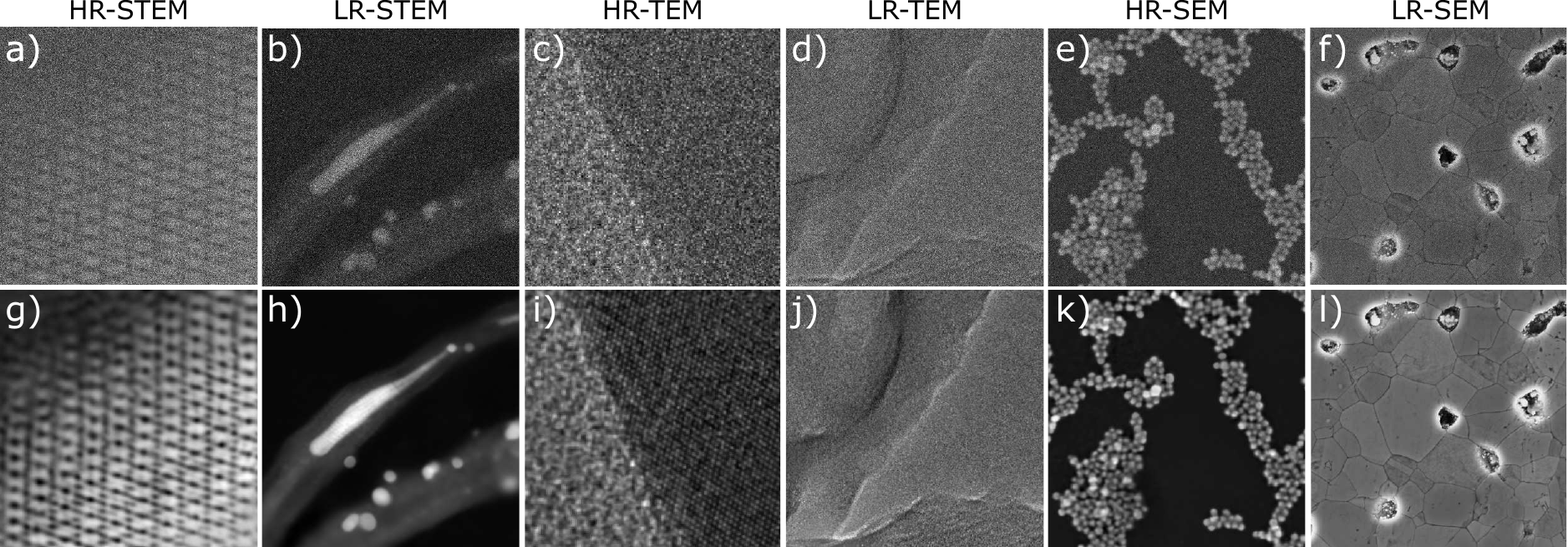}
	\caption{Experimental image restoration for various microscopy modalities. The top row illustrates the raw experimental images, while the bottom row displays the restored versions. Images (a), (b), (c), and (d) were obtained from reference \cite{Ede2020}, and images (e) and (f) were sourced from reference \cite{Aversa2018}.}
	\label{fig:em_restoration}
\end{figure}

One of the main advantages of our image restoration method is that the training data is generated using realistic physical models of the noise found in various microscopy modalities, as well as for an appropriate range of values for the noise model parameters, as detailed in the "Methods" section.  This methodology allows for the direct application of our network to experimental data, without requiring additional training for a particular specimen or microscope settings. Figure \ref{fig:em_restoration} illustrates the effectiveness of our approach on diverse types of random experimental microscopy images. The top row of this figure shows raw experimental images for HR-STEM, LR-STEM, HR-TEM, LR-TEM, HR-SEM, and LR-SEM. The bottom row shows the corresponding restored versions of these images.
These results show that the trained networks have excellent performance on experimental data and can effectively handle a wide range of microscopy images with varying resolution and noise levels. It is important to note that in this study, "high resolution" refers to images with round and symmetrical features, while "low resolution" refers to images with a variety of different features. Additional examples of restored experimental images for each microscopy modality can be found in the github repository \url{https://github.com/Ivanlh20/r_em}.\\

\begin{figure}[!ht]
	\centering \includegraphics[width=\linewidth]{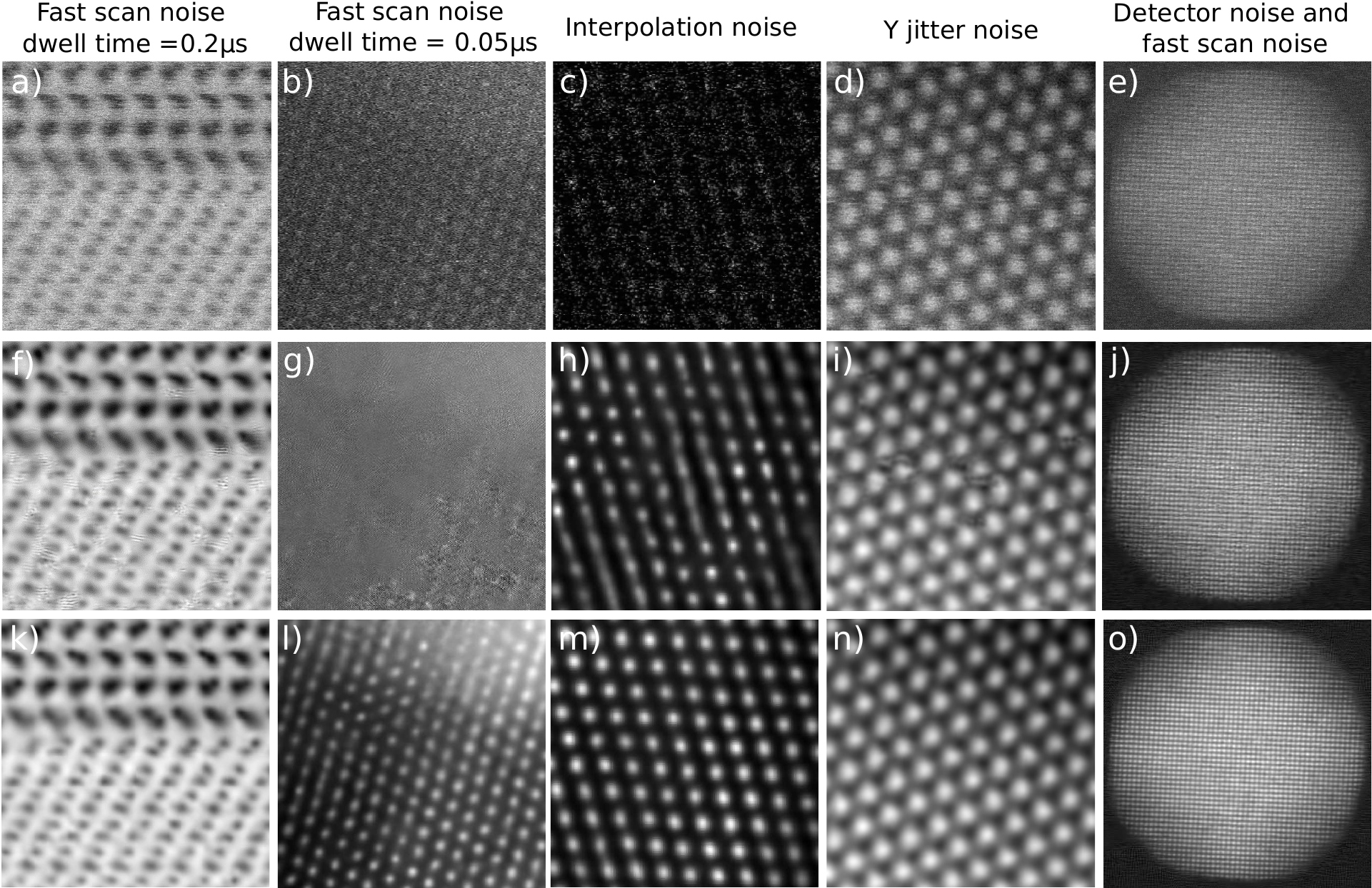}
	\caption{Raw STEM images alongside the results of a restoration process employing inaccurate and accurate models of the noise. The top row shows the original STEM images, while the second and third rows show the restored versions of the images trained with distorted data based on inaccurate and accurate noise models, respectively. Images (a)-(c) were obtained from our experimental datasets, whereas (d) and (e) were obtained from references \cite{Amini2018} and \cite{Altantzis2019}, respectively.}
	\label{fig:wrong_exp_modelling}
\end{figure}

The importance of using realistic physical models of the noise to generate distorted data, along with selecting the correct range of values for the noise model parameters, is demonstrated in Figure \ref{fig:wrong_exp_modelling}. This figure illustrates how these factors can impact the accuracy of the restored image. Figures \ref{fig:wrong_exp_modelling} (a) and (b) show two experimental STEM images that were acquired using a Fei $\hbox{Titan}^{3TM}$ S/TEM microscope. The images were obtained using fast scanning with dwell times of $0.2\mu s$ and $0.05\mu s$, respectively. The importance of accurately modelling fast scan distortion is evident from figures \ref{fig:wrong_exp_modelling} (f) and (g). In these figures, our network architecture was trained using a model, which was not sufficient to completely compensate for the spread of pixel intensities along the scanning direction (see Equation \ref{eq_fsd_stem_det} in the "S(T)EM noise model" section). If the dwell time decreases, these image artifacts become more pronounced, as shown in figure \ref{fig:wrong_exp_modelling} (g). While the manufacturer recommends using dwell times larger than $0.5\mu s$ to avoid image artifacts, correctly modelling fast scan distortion allows us to fully compensate for these artifacts, as shown in figures \ref{fig:wrong_exp_modelling} (k) and (l). The study of beam-sensitive materials and dynamic imaging will greatly benefit from the compensation of this distortion. Figure \ref{fig:wrong_exp_modelling} (c) shows a registered STEM image that contains interpolation noise. The interpolation process changes the dominant noise distribution, which can impact the restoration process, especially at low doses, as shown in Figure \ref{fig:wrong_exp_modelling} (h) where some atomic columns appear blurred. However, this issue can be addressed by including this type of noise in the training dataset, as explained in the "Methods" section. The effect of including this noise in the training dataset on the restored image can be seen in figure \ref{fig:wrong_exp_modelling} m), where all atomic columns become clearly visible. 
Figure \ref{fig:wrong_exp_modelling} (d) exhibits a STEM image with strong Y-jitter distortion. The impact of an incorrect range of values for this distortion during data generation on the restored image can be seen in figure \ref{fig:wrong_exp_modelling} (i), where some atomic columns appear split. After retraining the data with newly generated data containing the proper range of Y-jitter distortion, the neural network can correctly compensate for this image artifact, as shown in figure \ref{fig:wrong_exp_modelling} (n). In Figure \ref{fig:wrong_exp_modelling} (e), an experimental STEM image of a nanoparticle taken using a gas cell holder is shown \cite{Altantzis2019}. The dominant sources of noise in this image are detector noise and fast scan noise. Figure \ref{fig:wrong_exp_modelling} (j) shows a restored STEM image produced by our network architecture that was trained using a dataset generated with Poisson noise as the only source of STEM detector noise (as described by Equation \ref{eq_stem_det_noise_P} in the "S(T)EM noise model" section). However, this restored image exhibits strong artifacts despite using an accurate model for fast scan noise (as described by Equation \ref{eq_fsd_full} in the "S(T)EM noise model" section). After retraining our network architecture with a new dataset that includes the correct STEM detector noise (as described by Equation \ref{eq_stem_det_noise_F} in the "S(T)EM noise model" section), the restored image in Figure \ref{fig:wrong_exp_modelling} (o) shows a significant reduction in artifacts. Nonetheless, it is worth mentioning that some of the remaining artifacts in the image could be attributed to other sources of distortion not accounted for in our data modelling, such as the gas holder effect, charging artifacts, and residual electronic noise.

\begin{figure}[!ht]
    \centering
    \includegraphics[width=\linewidth]{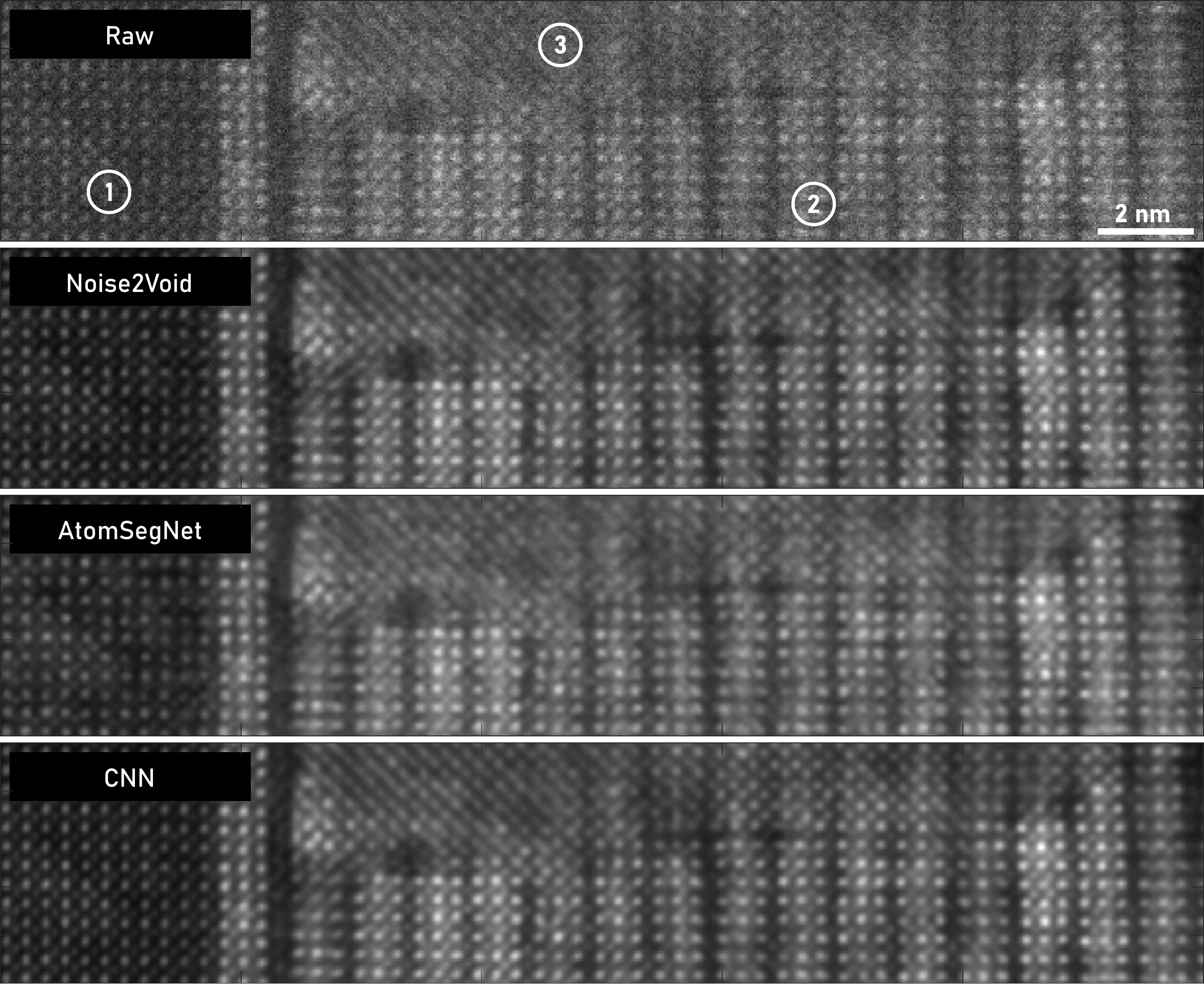}
    \caption[Comparison of different CNN-restorations]{Comparison of different CNN-restoration approaches on an experimental HAADF-STEM dataset of a $ BaHfO_3 $ nanoparticle (\circnum{3}) embedded in a superconducting $ REBa_2Cu_3O_{7-\delta} $ (REBCO)  matrix (\circnum{2}), which was epitaxially grown on a $ SrTiO_3 $ substrate(\circnum{1}). Images were acquired on a non-probe-corrected Titan microscope with 300~keV at KIT Karlsruhe. The data is descibed in detail in references \cite{Cayado2022} and \cite{Gruenewald2022}}. 
    \label{fig:concl_n2v}
\end{figure}

Another example that highlights the importance of properly modeling noise and distortion sources can be seen in Figure \ref{fig:concl_n2v}. In this figure, we compare the reconstruction performance of our CNN, AtomSegNet \cite{Lin2021}, and Noise2Void-NN (N2V) \cite{Krull2019}, which was retrained on the presented experimental image itself. The sample is a $BaHfO_3$ nanoparticle (figure \ref{fig:concl_n2v}-\circnum{3}) embedded in a superconducting $ REBa_2Cu_3O_{7-\delta} $ (REBCO)  matrix\cite{Cayado2022, Gruenewald2022} (figure \ref{fig:concl_n2v}-\circnum{2}), which was grown on a $SrTiO_3$ substrate  (figure \ref{fig:concl_n2v}-\circnum{1}). 
While all three networks successfully remove the noise from the image, there are notable differences in the reconstruction results. In region \circnum{1}, the N2V reconstruction recovers all the weaker intensities of the $ Ti+O $ columns to some degree, which is not the case for the AtomSegNet reconstruction. There, some of the columns blur or even disappear. Our CNN reliably recovers all atomic columns with superior contrast to the other two methods. Similar improvements are evident also in region \circnum{2} but most notably in region \circnum{3}. This region at the top of the image is also degraded, presumably by either FIB damage or carbon contamination. In both N2V and AtomSegNet reconstructions, features tend to blur into diagonal streaks, while our CNN recovers clearly distinguishable atomic columns and, given that the $ BaHfO_3 $ nanoparticle grew epitaxially on the $ SrTiO_3 $ substrate, that is indeed what would be expected \cite{Molina-Luna2015}. Considering the N2V network is a generic denoising network, the results are quite remarkable, albeit the additional training step is somewhat inconvenient from a user perspective. 
However, this example illustrates that the CNN presented in this work does not only benefit from the latest advances in deep learning, but also from the development of accurate, physically meaningful models of all distortions specific to HAADF-STEM. This CNN is shown to be accurate, not only in perceived contrast enhancement, but also in a quantitative way which boosts the accuracy and precision of atomic structure determination in ADF-STEM studies. 

\section*{METHODS}\label{sec:methods}
In single-shot EM image restoration, the goal is to estimate an undistorted image $y$ from a distorted image $x$. To achieve this, we train a generator $G$ using a deep neural network approach, which learns to estimate the corresponding undistorted image $y$ for a given input $x$. During the training procedure, a loss function is minimised to evaluate the quality of the results.

Traditionally, pixel-wise losses such as $\LX_{1}$ or $\LX_{2}$ have been used to obtain quantitative results for the image restoration problem \cite{Zhao2017}. However, these losses often lead to blurred images that do not look realistic. To address this, we propose a conditional generative adversarial network (cGAN) that trains both a generator and a discriminator. The generator $G$ maps the distorted image $x$ to the undistorted image $y_g=G(x)$, and the discriminator is trained to differentiate between real and generated images \cite{Isola2016}. We use pixel-wise losses to ensure quantitative results while restricting the GAN discriminator to model high-frequency details, resulting in sharper and more realistic restored images.

Our training is supervised, which requires input pairs of distorted and undistorted EM images. However, in practice, we only have access to distorted EM data. To overcome this, we can partially address the problem by collecting time-series EM images and using an average procedure based on rigid and non-rigid registration to generate an undistorted image. However, the combination of high-speed scans, jitter, and low-dose leads to highly correlated distortions \cite{BRGBS10}. Furthermore, long exposure to the electron beam can result in charging, beam damage, atom hopping and rotation of the specimen under study, which can further hamper the average procedure. Therefore, the only solution is to train the GAN using synthetic pairs of undistorted/distorted EM images.

\subsection*{Network architecture}
A GAN \cite{Goodfellow} is a powerful framework that encourages predictions to be realistic and thus to be close to the undistorted data distribution. A GAN consists of a generator (G) and discriminator (D) playing an adversarial game. A generator learns to produce output that looks realistic to the discriminator, while a discriminator learns to distinguish between real and generated data. The models are trained together in an adversarial manner such that improvements in the discriminator come at the cost of a reduced capability of the generator and vice versa. The GAN involves the generation of conditional data, which is fed to the generator and/or the discriminator \cite{Mirza2014}. The generator and discriminator architectures proposed here are adapted from those described in \cite{Zhang2018_1} and \cite{Isola2016}, respectively. The details of these architectures are discussed in the following sections.

\begin{figure}[!ht]
	\centering \includegraphics[width=\linewidth]{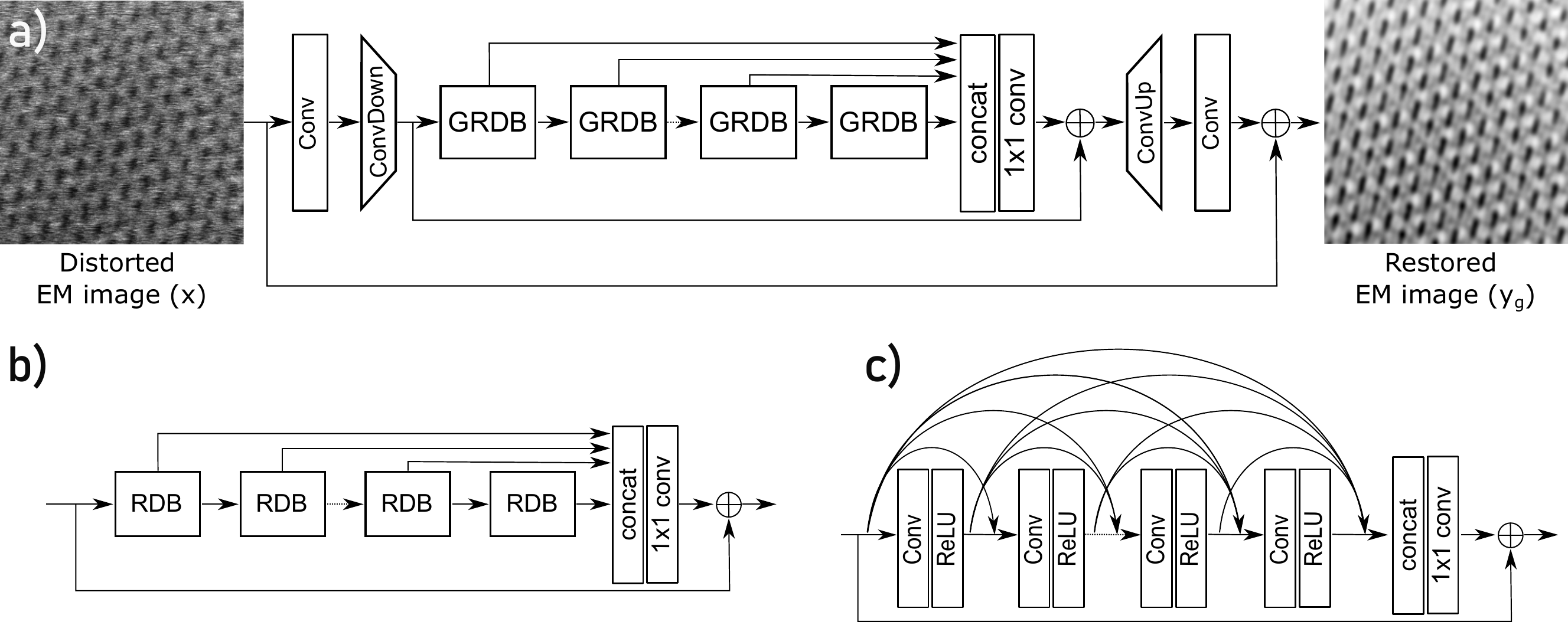}
	\caption{Concatenated Grouped Residual Dense Network (CGRDN) architecture for EM image restoration. (a) Overall architecture, (b) GRDB architecture used in (a), (c) RDB architecture used in (b).}
	\label{fig:generator_arch}
\end{figure}

\subheader{Generator architecture}\label{sec:gen_architecture}
Our generator architecture, called Concatenated Grouped Residual Dense Network (CGRDN), is shown in Fig. \ref{fig:generator_arch}. This network architecture is an extension of the GRDN for image denoising \cite{Kim2019}, which was ranked first for real image denoising in terms of the PSNR and the structure similarity index measure in the NTIRE2019 Image Denoising Challenge \cite{Ignatov2019}. The GRDB architecture is shown in Fig. \ref{fig:generator_arch}(a). The building module of this architecture is the residual dense block (RDB) \cite{Zhang2018_1}, which is shown in Fig. \ref{fig:generator_arch}(b). The original GRDN architecture can be conceptually divided into three parts. The first part consists of a convolutional layer followed by a downsampling layer based on a convolutional stride, the middle part is built by cascading GRDBs and the last part consists of an upsampling layer based on transposed convolution followed by a convolutional block attention module (CBAM) \cite{Woo2018} and a convolutional layer. The GRDN also includes the global residual connection between the input and the last convolutional layer. In the original version of the GRDN \cite{Kim2019}, residual connections are applied in three different levels (global residual connection, semi-global residual connection in GRDB, and local residual connection in each RDB). However, in the version submitted for the NTIRE2019 Image Denoising Challenge \cite{Ignatov2019}, residual connections for every 2 GRDBs were included.

Although it has been demonstrated that one architecture developed for a certain image restoration task also performs well for other restoration tasks \cite{Zhang2018_1, Plotz2018, Isola2016, Zhang2017}, an architecture for a given task will be data dependent. When applied to EM data, we found out that 2 modifications of GRDN are necessary in order to best handle the nature of our data, which involves different types and levels of distortions with high correlation between pixels:
\begin{enumerate}
	\item The cascading of the GRDN is replaced by feature concatenation, feature fusion, and a semiglobal residual connection. This allows us to exploit hierarchical features in a global way, which is important for highly correlated pixels that extend over a large area of the image.
	\item The CBAM, which is included in \cite{Zhang2018_1} is removed from our network. The reason for this is the use of large image sizes (256x256) for training, which reduces its gain \cite{Kim2019}.
\end{enumerate}

\begin{figure}[!ht]
	\centering \includegraphics[width=\linewidth]{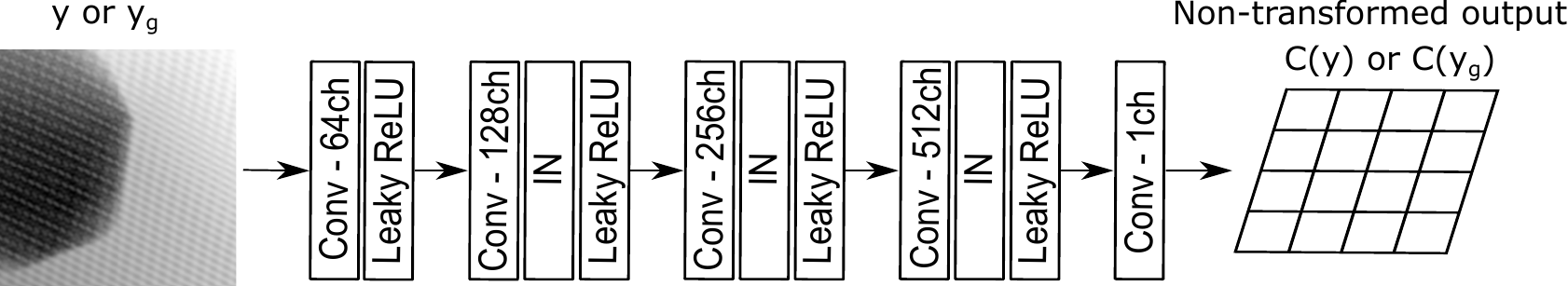}
	\caption{Patch discriminator architecture.}
	\label{fig_D}
\end{figure}

\subheader{Discriminator architecture}
The purpose of the discriminator network is to judge the quality of the output data resulting from the generator network. For our discriminator, we use the $70x70$ convolutional patch discriminator described in \cite{Isola2016} with some minor modifications. The zero-padding layers were removed and batch normalization layers \cite{Ioffe2015} were replaced by instance normalization layers (IN) \cite{Li2015}. Figure \ref{fig_D} shows the structure of the discriminator network. The result of the network is the non-transformed output $C(y)$ or $C(y_g)$ of dimensions $32x32$.
Some benefits of the discriminator architecture shown in Fig. \ref{fig_D} include that it is fully convolutional and it only penalizes structure at the scale of image patches. Furthermore, we enhance our discriminator based on the relativistic GAN, which has been shown to improve the data quality and stability of GANs at no computational cost \cite{Jolicoeur-Martineau2019}. Different from the standard discriminator, which estimates the probability that input data is real, a relativistic discriminator predicts the probability that real data $y$ is relatively more realistic than generated data $y_g=G(x)$. If we denote our relativistic average patch discriminator as $D_{Rap}(x)$, then the output of the discriminator can be written as:
\begin{eqnarray}
	D_{Rap}\left(y, y_g\right)=&\sigma\left(C(y)-\EXP[y_g]{C(y_g)}\right)\label{'eqn_drap_yyg'}\\
    D_{Rap}\left(y_g, y\right)=&\sigma\left(C(y_g)-\EXP[y]{C(y)}\right)\label{'eqn_drap_ygy'}
\end{eqnarray}
where $\sigma$ is the sigmoid function and $\EXP[x_1,...x_n]{.}$ is an operator representing the expectation value computed on the variables $x_1,...x_n$. In the next section, these functions will be used in the definition of the loss functions.

\subsection*{Loss function}
The loss function is the effective driver of the network's learning. Its goal is to map a set of parameter values of the network onto a scalar value, which allows candidate solutions to be ranked and compared. In our case, the discriminator and adversarial losses are based on the relativistic average GAN loss defined in \cite{Jolicoeur-Martineau2019}. We design our generator loss function as a sum of different contributions in such a manner that it keeps the quantitative information of the image at the pixel level and produces perceptually correct and realistic images. The different contributions of these loss functions are described in the following sections.

\subheader{$\LX_1$ loss}
Pixel-wise losses are advantageous to keep quantitative information of the ground truth image. In this work, we used the $\LX_{1}$ loss, which as compared to the $\LX_{2}$ loss yields less blurred results \cite{Zhao2017}. The $\LX_1$ loss can be written as:
\begin{eqnarray}
	\LX_{1} &=& \EXP[y, y_g]{w_y\NX{y-y_g}},\label{eq_loss_l1}\\
	w_y &=& 1/\max{\left(\sigma_{\min}, \SXP[y]{y}\right)}
\end{eqnarray}
where $w_y$ is a weighting factor that gives equal importance to each example regardless of its contrast, $\sigma_{\min}$ is a small value to limit the maximum scaling factor, and $\SXP[x_1,...x_n]{.}$ is an operator that represents the standard deviation calculated on the variables $x_1,...x_n$.

\subheader{$\LX_2$ loss}
Due to the design of our architecture, which is learning the residual difference between the distorted and undistorted image and based on the fact that distorted images can have few outliers in the distribution of pixel intensities (i.e. X-rays hitting the EM detector, saturation of the detector, low dose and dead-pixels), the output of the generator will show a strong correlation at those pixel positions. For this reason, we also used the $\LX_{2}$ loss which strongly penalized the outliers:
\begin{equation}
	\LX_{2} = \EXP[y, y_g]{w_y\NX{y-y_g}^2}
	\label{eq_loss_l2}
\end{equation}

\begin{figure}[!ht]
	\centering \includegraphics[width=\linewidth]{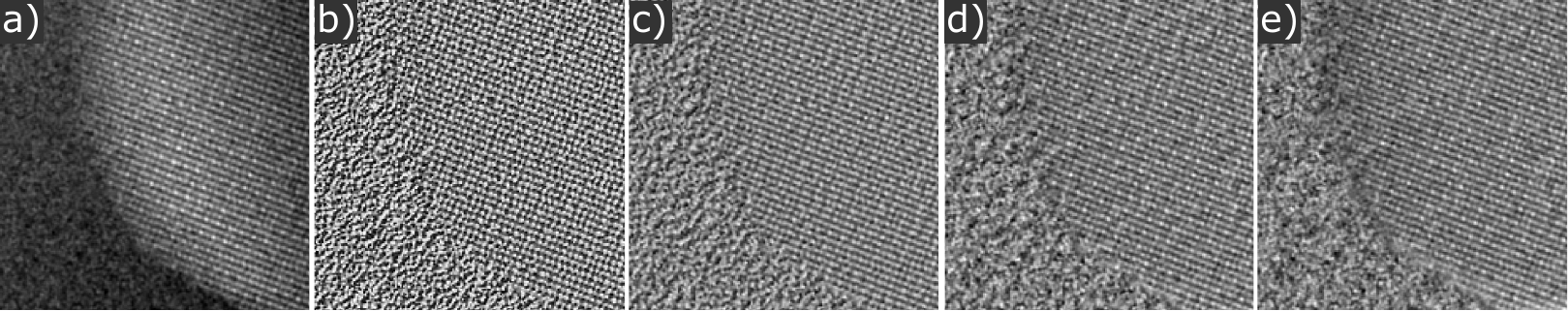}
	\caption{a) Undistorted ADF STEM image of a nanoparticle on a carbon support. Images are generated by applying the whitening transform to (a) by using different window sizes of (b) 2, (c) 4, (d) 8 and (e) 16.}
	\label{fig_lwt}
\end{figure}

\subheader{Multi-local whitening transform loss}
Local contrast normalisation (LCN) is a method that normalises the image on local patches on a pixel basis \cite{Jarrett}. A special case of this method is the whitening transform which is obtained by subtracting the mean and dividing by the standard deviation of a neighborhood from a particular pixel:
\begin{equation}
	y_{ij}^{S} = \left(y_{ij} - \EXP[\hat{S}]{y_{i, j}}\right)/\max{\left(\sigma_{\min}, \SXP[\hat{S}]{y_{i,j}}\right)},
\end{equation}
where $\hat{S}$ is a local neighbourhood around the pixel $i, j$ of window size $S$. The whitening transform makes the image patches less correlated with each other and can highlight image features that were hidden in the raw image due to its low local contrast. This effect can be seen in Fig. \ref{fig_lwt}a), which shows a simulated ADF-STEM image of a random nanoparticle on a carbon support. The edge of the nanoparticle shows low contrast due to its reduced thickness, resulting in lower intensity values. Based on this observation, we introduce a multi-local whitening transform (MLWT) loss which pays more attention to fine details independent of the intensity value. Specifically, the generated and the ground truth image are local whitening transforms corresponding to different window sizes of $2x2$, $4x4$, $8x8$, and $16x16$ pixels.

Using different windows sizes for the calculation of the whitening transform, we ensure that the relevant features present in the image are highlighted independently of its pixel size. Figs. \ref{fig_lwt}(b)-(e) show an enhancement of the edge of the nanoparticle as well as the carbon support after applying the whitening transform to Fig. \ref{fig_lwt}(a) by using different window sizes.\\

Then, we calculate the average $\LX_1$ loss for these 4 images:
\begin{equation}
	\LX_{mlwt} = \frac{1}{4}\sum_{S = 2, 4, 8, 16}\EXP[y^S, y_g^S]{\NX{y^S-y_g^S}}.
	\label{eq_loss_mlwt}
\end{equation}

\subheader{Fourier space loss}
In electron microscopy, Fourier space contains crucial information about the sample and any distortions that may be difficult to discern in real space. To address this issue, we introduce the $\mathcal{L}{\gamma}$ loss in the 2D Fourier transform of the difference between the generated data $y_g$ and the ground truth image $y$.Nevertheless, it is noted that high-frequency information typically possesses smaller values than low-frequency information. Consequently, to accentuate the high-frequency information, we apply a power transform to the aforementioned difference and define the loss function as follows:

\begin{equation}
\mathcal{L}{\text{fs-}\gamma} = \mathbb{E}{y, y_g}\left[|\mathcal{F}(y-y_g)|^{\gamma}\right],
\label{eq_loss_fs}
\end{equation}

Here, $\mathcal{F}$ symbolises the 2D Fourier transform, and $\gamma$ is a parameter in the range $(0.0, 1.0]$. In our investigation, we utilise $\gamma=0.125$.

\subheader{Constraint losses}
Some important parameters for EM quantification are the total intensity and the standard deviation of the images. The reason for this is that they carry information about physical quantities of the sample or microscope, such as the number of atoms, defocus and spatial and temporal incoherence \cite{VanAert2011, Beyer2016}. Therefore, we encourage that the restored images have to minimize the above quantities, resulting in the following two loss functions:
\begin{eqnarray}
	\LX_{mean} &=& \NX{\EXP[y]{y}-\EXP[y_g]{y_g}},\\
	\LX_{std} &=& \NX{\SXP[y]{y}-\SXP[y_g]{y_g}}.
\end{eqnarray}

\subheader{Adversarial loss}
The job of the relativistic adversarial loss is to fool the discriminator which can be expressed as:
\begin{equation}
	\LX_{Adv} = -\EXP[x, y]{\log \left(1-D_{Rap}(y, y_g) \right)} - \EXP[y_g]{\log\left(D_{Rap}(y_g, y)\right)},
\end{equation}
with $D_{Rap}(y, y_g)$ and $D_{Rap}(y_g, y)$ defined in equations \ref{'eqn_drap_yyg'} and \ref{'eqn_drap_ygy'}, respectively. This definition is based on the binary cross entropy between the ground truth and the generated images. Different from the conventional adversarial loss, in which $y$ is not used, our generator benefits from $y$ and $y_g$ in the adversarial training.

\subheader{Generator loss}
Our total generator loss function can be written as:
\begin{eqnarray}
	\LX_{G} &=& \LX_{pixel-wise} + \lambda_{Adv} \LX_{Adv},
	\label{eq_overall_loss}\\
	\LX_{pixel-wise} &=& \lambda_{1} \LX_{1} +  \lambda_{2} \LX_{2} + \lambda_{mlwt} \LX_{mlwt} + \lambda_{fs-\gamma} \LX_{fs-\gamma} + \lambda_{mean} \LX_{mean} + \lambda_{std} \LX_{std},
	\label{eq_pixel_wise_loss}
\end{eqnarray}
where $\LX_{pixel-wise}$ is our pixel-wise loss function, $\lambda_{1}$, $\lambda_{2}$, $\lambda_{mlwt}$, $\lambda_{fs-\gamma}$, $\lambda_{mean}$, $\lambda_{std}$ and $\lambda_{Adv}$  are the weighting parameters to balance the different loss terms.

\subheader{Discriminator loss}
Symmetrically to the relativistic adversarial loss, the relativistic discriminator is trying to predict the probability that real data is relatively more realistic than generated data, and it can be expressed as:
\begin{equation}
	\LX_{D} = -\EXP[x, y]{\log \left(D_{cRap}(x, y, y_g) \right)} - \EXP[x, y_g]{\log\left(1-D_{cRap}(x, y_g, y)\right)}.
\end{equation}
\subsection*{Data generation}
While it is possible to fully describe the electron-specimen interaction and image formation in an electron microscope, generating realistic EM image simulations for specimens on a support with sizes of a few nanometers is too time-consuming even with the most powerful GPU implementations of the multislice method \cite{Lobato2015, Lobato2016}. However, our goal is to train a neural network to correct EM distortions without the need to know the specific specimen or microscope settings. Therefore, we only need to generate undistorted images that closely mimic the appearance of real EM data, while the EM distortions must be accurately modelled. The generated undistorted images should also include physical parameters of the specimen and microscope settings, such as atomic sizes, atomic distances, atomic vibrations, lattice parameters, and relative intensities of atomic species, as well as acceleration voltage, aberrations, magnification, detector sensitivity, detector angles, and the transfer function of the detection system.

\subsubsection*{Specimen generation}
In order to optimise the simulation process, we generate a specimen that fully covers the extended simulated box size $\hat{l}_{xyz}^e$. This is an expanded version of the required simulation box size $\hat{l}_{xyz}$. The calculation of $\hat{l}_{xyz}$ starts by randomly selecting a pixel size $\dr$ within the range $[0.025, 0.90]\text{\AA}$. By using the required image size $(n_x, n_y)$, $n_z=\max(n_x, n_y)$ and $\dr$, the required simulation box size can be expressed as $\hat{l}_{xyz}=\{n_x\dr, n_y\dr, n_z\dr\}$. From these values, an extended number of pixels $n_i^e=n_i+round(d_{ext}/dr)$ and an extended simulation box size $\hat{l}_{xyz}^e=\{n_x^e\dr, n_y^e\dr, n_z^e\dr\}$ are obtained, where $d_{ext}$ is the maximum correlation distance for a given value of scanning distortions. The specimen generation is divided in 3 steps.

The first step of specimen generation involves randomly selecting a specimen type from the following options: crystalline specimen, amorphous specimen, or individual points. If the selected specimen is crystalline, the generation process starts by randomly choosing up to 16 unique atomic types with atomic number $Z$ in the range $[1, 103]$. The crystallographic space group is randomly chosen from a range $[1, 230]$. The lattice parameters and the angles of the chosen space group are selected randomly from a range $[3.1, 25.0]\text{\AA}$ and $[45^{\circ}, 120^{\circ}]$, respectively. Atomic positions of the asymmetric unit cells are generated randomly within the volume that is allowed by their space-group symmetry. This specimen generation process is subject to a physical constraint: after applying the space group symmetry to the atomic positions on the asymmetric unit cells, the minimum distance between the atoms in the unit cell must be within the range $[0.95, 7.0]\text{\AA}$. If this requirement is not met, the generation process is restarted.
The generation of amorphous specimens is based on randomly choosing only one atomic number $Z$ from the range $[1, 103]$. The atomic positions of amorphous specimens are generated by randomly placing atoms within the extended simulation box, subject to the requirement that the minimum distance between atoms is within the range $[0.95, 1.6]\text{\AA}$. This process continues until the desired density within the range $[2.0, 7.0] g/{cm}^3$ is achieved.
In contrast, the generation of individual points starts by randomly choosing a number of points within a given range of positive integers. The 3D positions of the particles are then generated randomly within the extended simulation box, subject to the requirement that the minimum distance between particles is within the range $[1, 20]dr$. This option is also used to generate low-resolution images.

The second step begins by randomly choosing between a specimen orientation along the zone axis or a random orientation. The probability of choosing a zone axis orientation is $0.75$. If the specimen is crystalline, the zone axis orientation is randomly chosen from the first eight main zone axes, and a small random mistilt angle is generated for the chosen orientation using a normally distributed random number with a standard deviation of $5^{\circ}$. For non-crystalline specimens, a random 3D orientation is generated. To prevent alignment of crystalline specimens along the $xy$ directions, an additional random rotation is applied along the $z$ axis.
For a given generated orientation, the specimen is oriented and cropped in the $xy$ plane so that it fits within the extended simulated box. This is followed by a random generation of a wedge on the specimen with a probability of $0.75$. The wedge can be generated on the top, bottom, or both surfaces of the specimen, each with a probability of occurrence of $0.33$. The wedge orientation is generated randomly in the $xy$ plane, and its angle is chosen randomly from the range $[5^{\circ}, 45^{\circ}]$.
Shapes can be applied to the specimen with a probability of $0.5$. To avoid any preference for the three different types of shapes, the probability of occurrence for each type is set to $0.33$. The first type of shape is a polygon rod, for which the number of cross-section vertices sliced along its length is randomly chosen from the range $[3, 15]$. The rod is also placed and oriented randomly. The radius of the polygon is chosen randomly from the range $[0.01, 0.5]\max(\hat{l}_{xyz})$. The second shape is a convex polyhedron, for which the radius and the number of vertices are chosen randomly from the ranges $[0.01, 0.5]\max(\hat{l}_{xyz})$ and $[4, 20]$, respectively. The third shape is a hard shape, in which all atoms on one side of a randomly generated $3d$ plane parallel to the $z$ orientation are removed. The application of a chosen shape can be used to either remove or keep the atoms of the specimen, with a probability of keeping the atoms of $0.5$.
Defects are generated randomly with a probability of $0.8$. The process starts by randomly selecting a number of atoms, $n_{sel}$, within the specimen. This number is chosen randomly from the range $[0, n_{max}]$, where $n_{max}$ is equal to the number of atoms in the specimen multiplied by $0.25$ and rounded to the nearest whole number. The positions of the selected atoms are randomly changed with a probability of $0.5$. This is done by adding a normally distributed random number with a standard deviation equal to the atomic radius to the position of each selected atom.

The final step of specimen generation adds a support layer with a probability of $0.95$. The support layer can be either crystalline or amorphous, each with a probability of $0.5$. The thickness of the support layer is chosen randomly from the range $[1, 30]\text{nm}$. The process described above for crystalline and amorphous specimen generation is used for the support layer, with the exception of shape generation. Finally, the generated atoms are added to the specimen.

\subsubsection*{Undistorted data generation}
\noindent \textbf{High/medium resolution electron microscopy data} can be synthesized as a linear superposition of the projected signal of each atom in the specimen at a given orientation. Moreover, each projected atomic signal can be modelled as a two-dimensional radial symmetric function, $f^i_Z(r)$, where the index $i$ refers to an atom with atomic number $Z$ in the specimen. Under this assumption, $y$ can be expressed as:
\begin{equation}
y = \sum_Z \sum_i f^i_Z(|\textbf{r}-\textbf{r}_i|),
\end{equation}
\noindent where $\textbf{r}$ is a two-dimensional vector. Additionally, we model $f_Z(r)$ for each atom with atomic number $Z$ as a weighted sum of Gaussian, Exponential, and Butterworth functions:
\begin{equation}
f_Z(r) = h_1 e^{-\frac{r^2}{2 \left(r^m_Z\right)^2}} + h_2 e^{-\frac{r}{r^m_Z}} + \frac{h_3}{1+(r/r^m_Z)^{2n}},
\label{eq_fz_mod}
\end{equation}
where $h_1$, $h_2$, $h_3$, $n$ and $r_m$ are the parameters of our model which are restricted to positive values. This parameterization has 3 benefits. First, it accurately models almost any simulated/experimental incoherent EM image. Second, it allows for an easy inclusion of physical constraints. Third, it only requires $5$ parameters. To allow realistic tails of $f_Z(r)$, we constrain $n$ to be a uniform random variable between $[4.0, 16.0]$. We would also like to emphasize that all numerical ranges for the data generation were fine-tuned based on analyzing around $2000$ real simulations of (S)TEM images for different specimens and microscope settings.\\
\noindent In order to encode physical information into this model, $r^m_Z$ is chosen proportionally to the transformed two-dimensional mean square radius, $\hat{r}_Z$, of the projected atomic potential, $V^p_Z(r)$ \cite{Lobato2014}:
\begin{equation}
	r^m_Z = a\times(\hat{r}_Z)^{\alpha} + b
\end{equation}
\noindent where
\begin{eqnarray}
	a &=& \SXP[Z]{\hat{r}_{Z}}/\SXP[Z]{(\hat{r}_{Z})^{\alpha}}, \\
	b &=& \EXP[Z]{\hat{r}_{Z}} - a\times\EXP[Z]{(\hat{r}_{Z})^{\alpha}},\\
	\hat{r}_Z &=& \left[\frac{\int_{0}^{\infty}r^2 V^p_Z(r)r\dr}{\int_{0}^{\infty} V^p_Z(r)r\dr}\right]^{1/2}
\end{eqnarray}
and $\alpha$ is a uniform random variable between $[0.75, 1.25]$. On the other hand, the linear coefficients $h_1$, $h_2$ and $h_3$ are randomly chosen within the range $[0.5, 1.0]$ with the following constraint:
\begin{equation}
	\int f_{Z_i}(r)dr > \int f_{Z_j}(r)dr, 	\text{ if } Z_i > Z_j
\end{equation}
where $Z_i$ and $Z_j$ are the atomic numbers of two elements of the specimen. This constraint arises from the fact that the integrated intensity of quasi-incoherently scattered electrons of a given atomic number is proportional to $Z^\gamma$, in which $\gamma$ is a real number between $1.0$ and $2.0$ depending on the microscope settings \cite{Hartel1996}.\\

The process of \textbf{generating low-resolution images} begins by randomly choosing a set of low-resolution image types from the following options: soft particles, sharp particles, grains, bands, boxes, and cracks. This stage uses the specimen type "individual points" to generate random positions where different objects will be placed. Finally, the low-resolution image is obtained by linearly superimposing these individual objects.

The generation of soft particles starts by randomly choosing a number of particles in the range $[15, 85]$. Each soft particle image is generated by randomly rotating the asymmetric version of Eq. \ref{eq_fz_mod}, where $r^m_Z = (r^{m_x}_Z, r^{m_y}_Z)$ and $r^{m_y}_Z = \alpha r^{m_x}_Z$, with $\alpha$ a random variable in the range $[0.8, 1.2]$.
In the case of sharp particles, there is a sharp transition between the border and background of the particle, and the particle can be either polygonal or elliptical with equal probabilities of occurrence. The process starts by randomly choosing a number of particles in the range $[15, 40]$. For the polygon option, the number of vertices is randomly chosen in the range $[3, 5]$. Each sharp particle image is generated by masking a 3D random positive plane intensity with its randomly rotated shape. This masking creates an intensity gradient over the $x-y$ plane such that the object does not appear flat.

Grain generation in $2D$ is performed using the Voronoi tessellation method \cite{Chew1985}, which is one of the available techniques for producing random polygonal grains within a domain. This process starts by randomly selecting a number of points within the range $[15, 175]$. Each grain image is created by masking a 3D random positive plane with its corresponding Voronoi cell. Additionally, the grain borderline is included with a probability of occurrence of $0.5$, where its intensity value is randomly assigned within the range $[0.5, 1.5]\times\textrm{mean}(\hbox{grain intensity})$.

EM images may exhibit contrast inversion related to the projected specimen, which can be easily simulated by inverting the image:
\begin{equation}
y \leftarrow \max(y) - \text{y}.
\end{equation}
The probability of this mechanism occurring was set to $0.5$. To introduce non-linear dependence between the generated image intensity and the projected specimen's structure, $y$ is non-linearly transformed with a probability of occurrence of $0.5$:
\begin{equation}
y \leftarrow |y|^{\beta}
\end{equation}
where $\beta$ is a uniform random number selected from the range $[0.5, 1.5]$.

To further break this linearity, a random background was added to $y$. The background is randomly chosen between a 3D plane and a Gaussian, with an occurrence probability of $0.5$ for each. In the first case, a randomly orientated positive 3D plane is generated with a random height between $[0, \max(y)/2]$. In the second case, the Gaussian centre and its standard deviation are randomly chosen within the range of the $xy$ simulation box size and $[0.2, 0.6]\times\min(n_x, n_y)$, respectively. From the analysis of the experimental and simulated data, we found that the ratio $r_{std/mean} = \SXP[]{\text{y}}/\EXP[]{\text{y}}$ is between $[0.01, 0.35]$. Therefore, if the EM image does not fulfill the latter constraint, then it is linearly transformed as:
\begin{equation}
	y \leftarrow cy + d
\end{equation}
\noindent where $c$ and $d$ are chosen to bring $r_{std/mean}$ within the range of the constraint. Finally, the EM image is normalized through dividing by its maximum value.
\begin{equation}
	y \leftarrow \frac{y}{\max(y)}
\end{equation}
Note that the correct parameterization of the model and the randomness of its parameters are subject to physical constraints allowing to encode information in the generated high/medium resolution EM image of the atomic size, atomic vibration, relative intensities between atomic species, detector angle, acceleration voltage, aberrations and/or detector sensitivity.

\subsubsection*{TEM noise model}
The TEM noise model is based on the fact that TEM images are recorded using parallel illumination, and that most signal acquisitions for electrons are set up so that the detector output is directly proportional to the time-averaged flux of electrons reaching the detector. In case of TEM, the electrons are detected indirectly using a charge coupled device (CCD) sensor \cite{Spence1988} or a complementary metal oxide semiconductor (CMOS) sensor \cite{Tietz2008}, or directly using a direct electron detector \cite{Clough2014}.

For indirect detection, primary electrons are converted to photons in a scintillator, which are then directed to the CCD/CMOS sensor through a lens or fiber optic coupling. In contrast, for direct electron detectors, the CMOS sensor is directly exposed to the electron beam.

\subheader{TEM camera modulation-transfer function}
\noindent Scattering of incident electrons over the detector leads to the detection of electrons in multiple pixels, which can be quantitatively described using the modulation-transfer function (MTF). Because the effect of the MTF is to produce an isotropic smear out of features on the recorded TEM image, which in general cannot be distinguished from an undistorted TEM image recorded with other microscope settings, we embedded this effect into the undistorted TEM image by convolving it with the point-spread function (PSF), which is the Fourier transform of the MTF:
\begin{equation}
y \leftarrow y\otimes\text{PSF}.
\end{equation}
\noindent The MTF itself can be separated into a rotationally symmetric part, $\text{MTF}_r$, describing the spread of electrons in the detector, and a part describing the convolution over the quadratic area of a single pixel. This yields the following equation:
\begin{equation}
\text{MTF} = \text{MTF}_r\operatorname{sinc}(\pi u/2)\operatorname{sinc}(\pi v/2),
\end{equation}
where the Fourier space coordinates $(u,v)$ are defined in units of the Nyquist frequency \cite{Thust2009}. Furthermore, we found that the general shape of $\text{MTF}_r$ can be expressed parametrically as:
\begin{equation}
\text{MTF}_r = a e^{-\frac{g^2}{2b^2}} + (1-a)e^{-\frac{g^2}{2c^2}},
\end{equation}
where $a$, $b$ and $c$ are positive real numbers. These numbers were randomly generated until they fulfill the constraint that on a numerical grid of 1000 points with a length of 10 units of the Nyquist frequency, the $\text{MTF}_r$ is a positive and monotonically decreasing function.

\subheader{TEM detector noise}
TEM detectors are subject to three main sources of noise: shot noise, dark current noise, and readout noise. These noise sources can be classified into two types: temporal and spatial noise. Temporal noise can be reduced by frame averaging, whereas spatial noise cannot. However, some spatial noise can be mitigated by using techniques such as frame subtraction or gain/offset correction. Examples of temporal noise discussed in this document include shot noise, reset noise, output amplifier noise, and dark current shot noise. Spatial noise sources include photoresponse non-uniformity and dark current non-uniformity. Each of these noise sources can lower the SNR of a sensor imaging device.

\subheader{Photon shot noise}
After the initial conversion of the incident electron to its photon counterpart, the generated photons will hit the photosensor pixel area, liberating photo-electrons proportional to the light intensity. Due to the quantum nature of light, there is an intrinsic uncertainty arising from random fluctuations when photons are collected by the photosensor. This uncertainty is described by the Poisson process $\mathbb{P}$ with mean $\alpha x$, where $\alpha$ is a dose scale factor.

The distribution of $\alpha$ is exponential, with a scale parameter of $0.5$ and a range $[0.5, 750]/\EX\{y\}$. The use of the exponential distribution yields higher probabilities for the generation of images at lower doses which is the focus of our research. The division by $\alpha$ in the equation below brings $x$ back to its original range:
\begin{equation}
x \leftarrow \frac{P(\alpha x)}{\alpha}
\label{eq_tem_sn}
\end{equation}

\subheader{Fixed-pattern noise}
Fixed-pattern noise (FPN) is a pixel gain mismatch caused by spatial variations in the thickness of the scintillator, fiber-optic coupling, substrate material, CCD bias pattern, and other artifacts that produce variations in the pixel-to-pixel sensitivity and/or distortions in the optical path to the CCD or in the CCD chip itself \cite{Vulovic2010}. Since FPN is a property of the sensor, it cannot be fully eliminated. However, it can be suppressed using a flat-field correction procedure. We model the remaining distortion as a normal distribution $\mathbb{N}$ with zero mean and standard deviation $\sigma_{fpn}$.
\begin{equation}
	x \leftarrow x + x \mathbb{N}(0, \sigma_{fpn})
	\label{eq_tem_fpn}
\end{equation}

\subheader{Dark-current noise}
Dark current is the result of imperfections or impurities in the depleted bulk Si or at the $SiO_2/Si$ interface.  These sites introduce electronic states in the forbidden gap which allows the valence electrons to jump into the conduction band and be collected in the sensor wells. This noise is independent of electron/photon-induced signal, but highly dependent on device temperature due to its thermal activation process \cite{Kodak2005}.

\subheader{Dark-current nonuniformity}
Dark-current nonuniformity (DCNU) arises from the fact that pixels in a hardware photosensor cannot be manufactured exactly the same and there will always be variations in the photo detector area that are spatially uncorrelated, surface defects at the $SiO_2/Si$ interface, and discrete randomly-distributed charge generation centers \cite{Konnik2014}. This means that different pixels produce different amounts of dark current. This manifests itself as a fixed-pattern exposure-dependent noise and can be modelled by superimposing two distributions. The Log-Normal distribution ($ln\mathbb{N}$) is used for the main body and the uniform ($\mathbb{U}$) distribution is used for the "hot pixels" or "outliers" \cite{Gow2007}.
\begin{equation}
	\textrm{DCNU} \leftarrow ln\mathbb{N}(\mu, \sigma)+ \mathbb{U}(a,b)
	\label{eq_tem_dcnu}
\end{equation}
with $\mu$ the mean value, $\sigma$ the standard deviation, $a= \mu + 5\sigma$, and $b= \mu + 8\sigma$. 

\subheader{Dark-current shot noise}
Additional noise arises from the random arrival of electrons generated as part of the dark signal, which is governed by the Poisson process. To simulate a single frame, it is necessary to apply shot noise to the DCNU array.
\begin{equation}
	x \leftarrow x +\mathbb{P}(\text{DCNU})
	\label{eq_tem_dcsn}
\end{equation}

\subheader{Readout noise}
Readout noise is temporal noise and is generally defined as the combination of the remainder circuitry noise sources between the photoreceptor and the ADC circuitry. This includes thermal noise, flicker noise and reset noise \cite{Irie2008}.

\subheader{Thermal noise}
Thermal noise arises from equilibrium fluctuations of an electric current inside an electrical conductor due to the random thermal motion of the charge carriers. It is independent of illumination and occurs regardless of any applied voltage. The noise is commonly referred to as Johnson noise, Johnson-Nyquist noise, or simply white noise. It can be modelled by the normal distribution with zero mean and an appropriate standard deviation $\sigma$ \cite{Irie2008}.
\begin{equation}
x \leftarrow x+ \mathbb{N}(0,\sigma)
\label{eq_tem_rotn}
\end{equation}

\subheader{Flicker noise}
Flicker noise, also known as $1/f$ noise or pink noise, is often caused by imperfect contacts between different materials at a junction, including metal-to-metal, metal-to-semiconductor, and semiconductor-to-semiconductor. MOSFETs are used in the construction of CMOS image sensors, which tend to exhibit higher levels of $1/f$ noise than CCD sensors \cite{Konnik2014}. The amount of flicker noise in a CCD sensor depends on the pixel sampling rate. The equation below describes the effect of flicker noise on a signal $x$:
\begin{equation}
x \leftarrow x + \FT(\mathbb{N}(0, \sigma)/f)
\label{eq_tem_rofn}
\end{equation}
Here, $\FT$ is the two-dimensional Fourier transform, $\sigma$ is the appropriate standard deviation, and $f$ is the reciprocal distance.

\subheader{Reset noise}
Before a measurement of the charge packet of each pixel is taken, the sense node capacitor of a specific row is reset to a reference voltage level. This causes all pixels in that row to be exposed to noise coming in through the reset line, transfer gate, or read transistor. As a result, images may have horizontal lines due to the fixed and temporal components of the noise. This type of noise, known as reset noise (RN), follows a normal distribution with mean zero and a standard deviation $\sigma$. It can be simulated by adding a random intensity value, generated for each row, to the intensity values of all pixels in that row \cite{Gow2007}:
\begin{equation}
x \leftarrow x+ \mathcal{N}(0, \sigma)
\label{eq_tem_dorn}
\end{equation}

\subheader{Black pixel noise}
Black pixels are dots or small clusters of pixels on the sensor that have significantly lower response than their neighbors, resulting in black spots on the image. Some black pixels may be created during the production process of the CCD camera, while others may appear during its lifetime. Black pixels are time-invariant and will always appear at the same locations on the image. They can be modelled by generating a sensitivity mask ($S_\text{Black}$) with a spatially uniform distribution of a specified number of black points. Regions can be generated by applying a random walk process for a given number of random steps to the black point positions.
The equation below describes the effect of black pixels on a signal $x$:
\begin{equation}
x \leftarrow x S_\text{Black}
\label{eq_tem_bpn}
\end{equation}

\subheader{Zinger noise}
Zingers are spurious white dots or regions that can appear randomly in CCD images \cite{ZINGER1961}. Electron-generated X-rays, cosmic rays, and muons can produce a burst of photons in the scintillator, resulting in white spots or streaks in the image. Radioactive elements (such as thorium) present in fiber-optic tapers can also cause zingers \cite{Vulovic2010}. They can be modelled by generating a sensitivity mask ($S_\text{Zinger}$) with a spatially uniform distribution of a specified number of zinger points. Similar to the black pixel noise, regions can be generated by applying a random walk process for a given number of steps to the zinger point positions:
\begin{equation}
x \leftarrow x S_\text{Zinger}
\label{eq_tem_zpn}
\end{equation}

\subheader{Upper-clip noise}
Upper clip noise, also known as saturation noise, is a type of noise that occurs when the intensity value of a pixel exceeds the maximum value that the CCD sensor can detect. This causes the pixel to be "clipped" at the maximum value, resulting in an overly bright image with lost details. This type of noise can be modelled by setting a threshold value for the maximum intensity and clipping any pixel values above that threshold $T_u$:
\begin{equation}
x \leftarrow \min(x, T_u)
\label{eq_tem_ucn}
\end{equation}

\begin{figure}[!ht]
\centering \includegraphics[width=\linewidth]{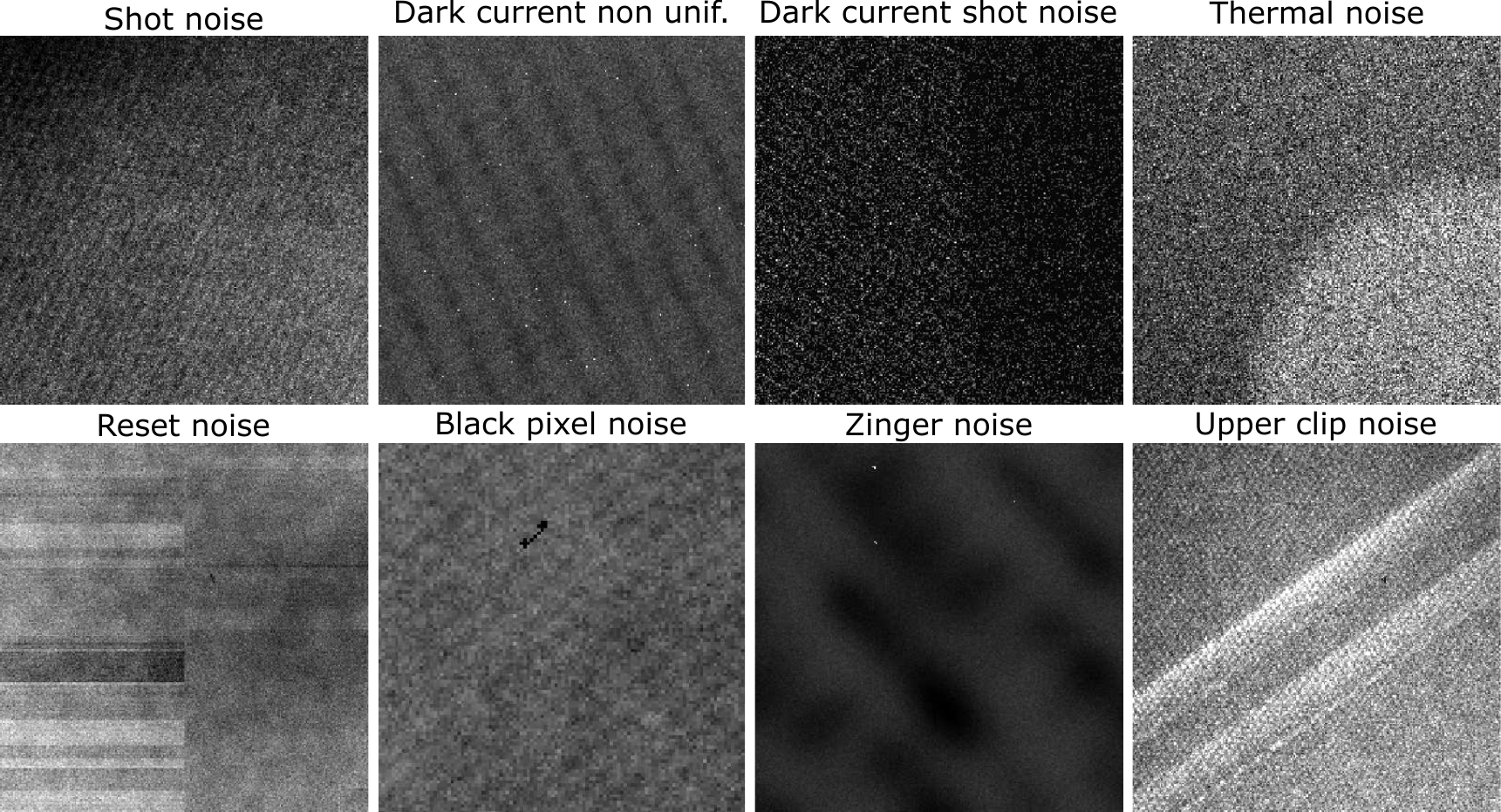}
\caption{Simulated TEM images with random distortions showing the various types of noise.}
\label{fig:tem_distortions}
\end{figure}

\subheader{Quantisation noise}
To generate a digital image, the analog voltage signal read out during the last stage is quantized into discrete values using analog-to-digital conversion (ADC). This process introduces quantization noise, which can be modelled with respect to the ADC gain $\alpha$:
\begin{equation}
x \leftarrow \text{round}(\alpha x)
\label{eq_tem_qtn}
\end{equation}
Figure \ref{fig:tem_distortions} shows simulated TEM images with different types of noise. These distortions have been randomly added to the images to mimic real TEM conditions and make it easier to identify the different types of noise.

\subsubsection*{S(T)EM noise model}
S(T)EM images are formed one pixel at a time by scanning a convergent electron beam along scan lines across the sample with constant stationary probing, which is known as dwell time. The dimension of each square-shaped pixel in the physical space is determined by the magnification. The scanning direction is called the fast/row scan direction. For conventional scan patterns, the scanning begins at the top left corner and after scanning one row of $n$ pixels, the electron probe moves to the next row's first pixel. The time required to move the beam to the beginning of the scan line is commonly known as fly-back-time. Inaccuracies in beam positions during the scanning process give rise to characteristic scan-line/jitter distortions. Despite all technical improvements in the design of high-performance S(T)EM \cite{KFBEA08}, the presence of these distortions on the recorded images still hampers the extraction of quantitative information from the sample under study \cite{Jones2017}.

\begin{figure}[!ht]
	\centering \includegraphics[width=\linewidth]{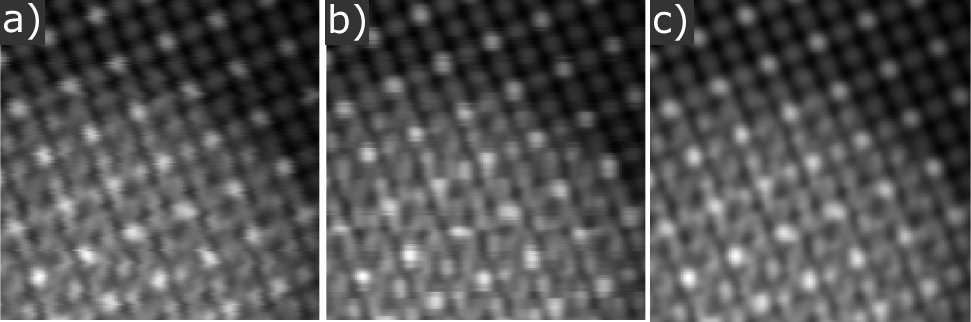}
	\caption{Image (a) and (b) are distorted jitter images along the fast and slow scan direction, respectively. (c) Undistorted ADF STEM image of a random sample.}
	\label{fig:jitter}
\end{figure}

\subheader{Scanning jitter distortion}
Scanning jitter is caused by beam instabilities while scanning a raster pattern across the sample during the image acquisition process. There are two distinguishable jitter effects: X-jitter causes random pixel shifts along the fast-scan direction, while Y-jitter causes stretching or squishing of scan lines or line interchanges along the slow-scan direction \cite{LJPN13}. Although these displacements are not completely random due to serial acquisition, they depend on the previous scan position. Realistic modelling of scanning jitter distortion can be achieved using the Yule-Walker correlation scheme on time series \cite{Yule1927, GW31}. Furthermore, the fast and slow scanning directions can be modelled independently due to their different time scales. Here, we focus on displacement series in discrete pixels, in which each term of the series depends on the previous one. Mathematically, these displacement series can be described as:
\begin{equation}
	\begin{split}
		\Delta^k_t &= \frac{a^k_t}{\sqrt{(1-\phi^2_t}} \quad \textrm{if } k=1\\
		\Delta^k_t &= \phi\Delta^{k-1}_t + a^k_t \quad \textrm{if } k>1
	\end{split}
	\label{eq_jitter}
\end{equation}
where $t={x, y}$ and $k$ is the pixel index along a given $t$ direction. $\phi_t$ is the correlation coefficient which describes the coupling between two consecutive values of the series within the range $[0, 1]$. $a^i_t$ is a normally distributed random number with zero mean and standard deviation $\sigma_t$. The distorted image is created by using bicubic interpolation and evaluating on the non-regular grid, which is built by adding the positions of the regular grid and the generated displacements.
\begin{equation}
x \leftarrow \text{SJ}(y)
\label{eq_stem_sjd}
\end{equation}
The described effects of individual jitter distortions for $\sigma_x=\sigma_y=0.75$ and $\phi_x=\phi_y=0.6$ along the fast and slow scan directions can be seen in Fig. \ref{fig:jitter}(a) and Fig. \ref{fig:jitter}(b), respectively. Fig. \ref{fig:jitter}(c) shows the undistorted ADF STEM random generated image.
Based on our analysis of experimental data, we set the occurrence probability of jitter distortion to $0.9$. In addition, we assign the occurrence probability of the X-jitter, Y-jitter and the XY-jitter to $0.25$, $0.25$ and $0.50$, respectively. The values of $\sigma_t$ and $\phi_t$ are randomly chosen within the range $[0.0025, 0.8]\text{\AA}$ and $[0.0, 0.7]$, respectively.

\subheader{S(T)EM detector noise}
Electrons are detected by a scintillator coupled to a photomultiplier tube (PMT) via a mirror or reflective tube. Impact of the incident electrons on the scintillator cause photons to be emitted, which are directed to the PMT through a light pipe. The PMT consists of a photocathode that emits photoelectrons when illuminated by these photons, followed by a series of stages amplifying the signal. The resulting current at the anode can be measured using conventional ADC electronics \cite{Ishikawa2014}. The statistics of the electron multiplication as a series of Poisson events with full width at half maximum (FWHM) of the pulse at the anode per single incident electron is given by \cite{Burle1984}:
\begin{equation}
\textrm{FWHM} = 2\sqrt{2 \log{2}}m_c \eta G \sqrt{\frac{1-\eta+\frac{1}{\delta-1}}{m_c\eta} + \frac{\delta_c^2}{m_c^2}}
\label{eq_stem_det_noise_FWHM}
\end{equation}
This equation assumes that the secondary gain $\delta$ at each stage inside the PMT is the same. In this equation, $G$ represents the PMT gain, $\eta$ is the detective quantum efficiency, $m_c$ is the number of photons collected per incident electron, and $\delta_c^2$ is the variance of that number \cite{Burle1984}.
A good approximation for the noise spectrum of a photomultiplier is the Poisson distribution, which can be approximated by a Gaussian distribution for large means. Since for each electron reaching the scintillator, around 100 photons reach the cathode of the photomultiplier, a Gaussian approximation can be used with standard deviation
\begin{equation}
	\sigma = m_c \eta  G \sqrt{\frac{1-\eta+\frac{1}{\delta-1}}{m_c\eta} + \frac{\delta_c^2}{m_c^2}}
	\label{eq_stem_det_noise_sigma}
\end{equation}
In addition, the number of electrons hitting the scintillator is described by the Poisson process ($\mathbb{P}$) \cite{Mittelberger2018}. The signal can therefore be constructed in two steps:
\begin{equation}
    x \leftarrow \mathbb{P}(\alpha x)
	\label{eq_stem_det_noise_P}
\end{equation}
\begin{equation}
    x \leftarrow (x + \mathbb{N}(0, \sigma))/\alpha 
	\label{eq_stem_det_noise_F}
\end{equation}
where $\alpha$ is a dose scale factor. Dividing by $\alpha$ in the latter equation brings $x$ back to approximately its original range.

\subheader{Fast scan noise}
Fast scan noise arises due to the use of short dwell times during data acquisition and appears as horizontal blur in the recorded images. This effect can also be seen in the Fourier domain as a damping effect on the high frequencies in the horizontal direction. This blurring is caused by the finite decay time of the detection system, which consists of a scintillator, a photomultiplier, and additional readout electronics \cite{Mittelberger2018, Mullarkey2021}. In addition to blurring in the horizontal direction, fast scans may introduce other artifacts due to the limited response time of the scan coils. In particular, strong distortions may appear on the left-hand side of the images due to the discontinuity in the scan pattern between consecutive lines. This can be avoided by using a small delay (flyback time) between scanning lines. The optimal value of this delay is hardware-specific, but results in additional dose to the sample, which will be localized on the left-hand side of each image \cite{KrivanekCD2008}.
In general, the effect of fast scan distortion can be modelled by convolution in one dimension along the fast-scan direction between $x$ and the point spread function (PSF) of the system. After careful analysis of the experimental data, we find that the PSF of the system can be decomposed into contributions from the detector and the readout system.
\begin{equation}
	\textrm{Im}_{fsd}(x, y) = \textrm{Im}\circledast \text{psf}_{detector} \circledast \text{psf}_{readout}
	\label{eq_fsd_full}
\end{equation}
with
\begin{equation}
	\text{psf}_{detector} = \left\{ \begin{array}{cc} \frac{\alpha}{4\pi^2x^2+\alpha^2} &: x <= 0 \\ 0 &: x>0 \end{array}\right.
	\label{eq_fsd_stem_det}
\end{equation}
\begin{equation}
	\text{psf}_{readout} = \left\{ \begin{array}{cc} a e^{-x/\beta}\sin(2\pi x/\gamma+\theta) &: x <= 0 \\ 0 &: x>0 \end{array}\right.
	\label{eq_fsd_readout}
\end{equation}

where 

\begin{equation}
	a = \frac{\beta \gamma \left (\gamma \sin(\theta) + 4\pi\beta cos(\theta)\right)}{\gamma^2 + 16\pi^2\beta^2}
\end{equation}

is the normalization factor which ensures that the total integral of the $\text{psf}_{readout}$ is equal to 1, $k$ is the pixel value in real space, and $\alpha$ is the parameter of the Lorentzian function that describes the PSF of the detector. The parameters $\beta$, $\gamma$, and $\theta$ are the parameters of the damped harmonic oscillator which is used to describe the PSF of the readout system. The model parameters were obtained by fitting to experimental images and by applying random variation to the fitting parameters.

\subheader{Row-line noise}
Row-line (RL) noise arises due to the non-response of the detector over some pixels during the scanning process along the fast-scan direction. This noise can be modelled by generating a random number of row lines with random length. The pixel intensities of the lines in the image are replaced by their average intensity multiplied by a random factor within the range $[0.5,1.5]$. This can be represented as:
\begin{equation}
x \leftarrow \mathbb{RL}(x)
\label{eq_stem_rld}
\end{equation}

\subheader{Black pixel noise}
Black pixels are randomly occurring pixels that have significantly lower values than their neighbouring pixels, causing black spots to appear in the image. These black pixels may result from information loss during data transmission, cosmic rays, or the detector's non-response. As black pixels are time-dependent, they can be modelled by generating a sensitivity mask ($S_\text{Black noise}$) with a spatially uniform distribution of a specified number of black points. This can be represented mathematically as:
\begin{equation}
x \leftarrow x S_\text{Black noise}
\label{eq_stem_bpn}
\end{equation}
However, in the case of SEM images, black spots in the images may be attributed to pores present in the sample, and hence, this type of distortion is not generated.

\subheader{Zinger noise}
Zingers are random white dots that appear in an image. They are caused by bursts of photons produced by electron-generated X-rays, cosmic rays, and muons in the scintillator \cite{Vulovic2010}. Zinger noise can be simulated by creating a sensitivity mask ($S_\text{Zinger noise}$) with a spatially uniform distribution of a specified number of Zinger points.
\begin{equation}
x \leftarrow x S_\text{Zinger noise}
\label{eq_stem_zpn}
\end{equation}

\subheader{Upper-clip noise}
Upper clip noise, also known as saturation noise, occurs when the intensity value of a pixel exceeds the maximum value that the analog-to-digital converter can detect. This causes the pixel to be "clipped" at the maximum value, resulting in an overly bright image with lost details. This type of noise can be modelled by setting a threshold value for the maximum intensity and clipping any pixel values above that threshold $T_u$.
\begin{equation}
x \leftarrow \min(x, T_u)
\label{eq_stem_ucn}
\end{equation}

\begin{figure}[!ht]
	\centering \includegraphics[width=\linewidth]{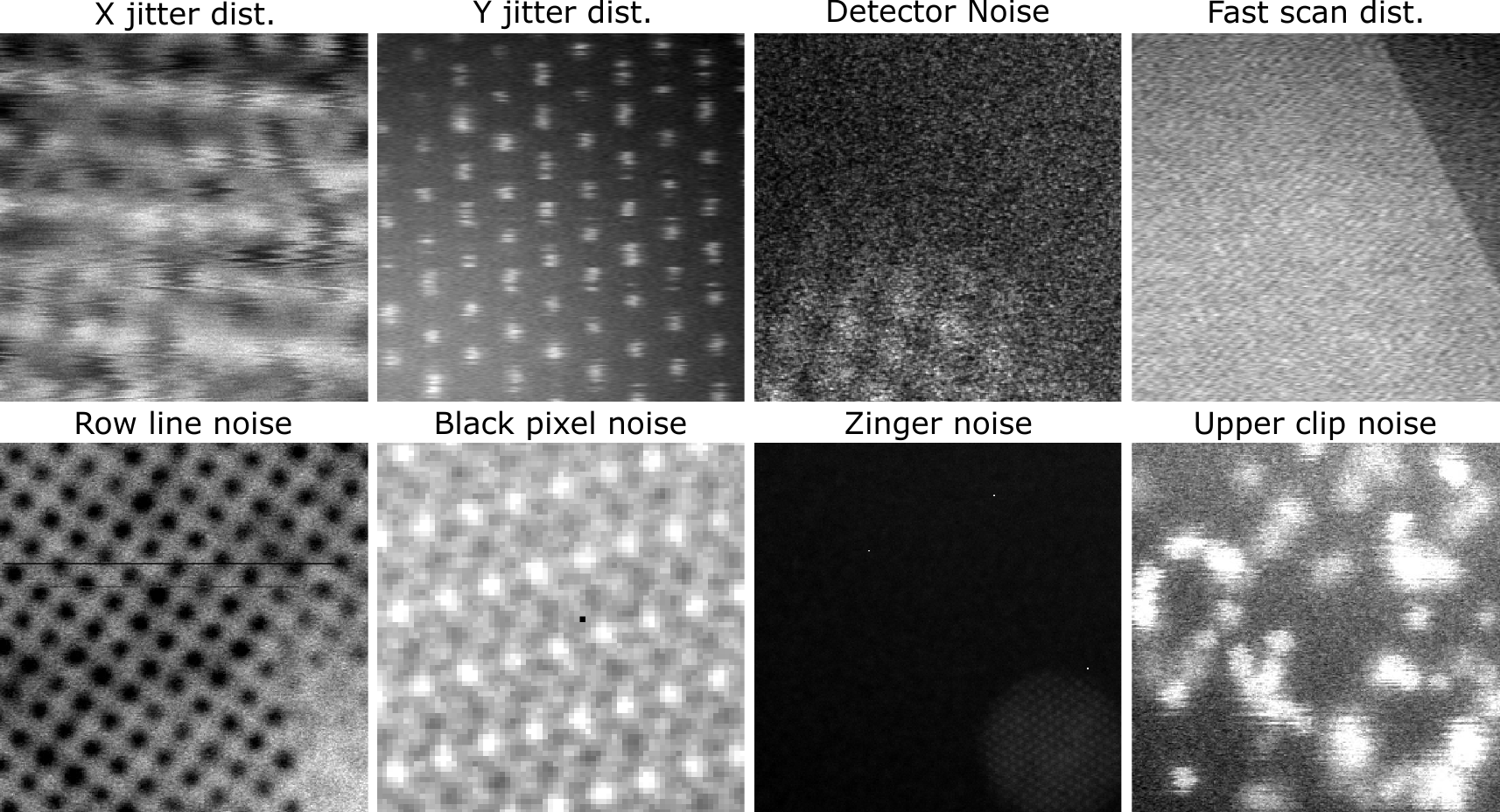}
	\caption{Random distorted simulated STEM images showing the various types of noise.}
	\label{fig:stem_distortions}
\end{figure}

\subheader{Quantisation noise}
To generate an image in digital form, the analog voltage signal read out during the last stage is quantized into discrete values using an ADC with a gain $\alpha$. This process introduces quantisation noise.
\begin{equation}
x \leftarrow round(\alpha x)
\label{eq_stem_qtn}
\end{equation}
Figure \ref{fig:stem_distortions} shows simulated STEM images of the different types of noise that can be found in STEM images. These distortions were randomly added to the images to simulate real STEM conditions and make it easier to identify the different types of noise. 

\subsubsection*{Post-processing distortions}
Post-processing distortions are typically added after the image is recorded. These distortions, such as interpolation and blurring, can affect the noise in the image in a non-linear way. Post-processing distortions can also include annotations and cropping, which replace part of the original image. Ideally, these distortions should be preserved by the restoration process.
\textbf{Interpolation distortions} may happen when a user applies a transformation function to the image before it is restored. This might be done to make the image suitable for further post-processing or to better visualise an area of interest. Interpolation distortion can be modelled by applying a random transformation, such as a random linear transformation matrix, to the training image pair.
\textbf{Gaussian blurring} is a way of distorting an image to reduce noise and improve the SNR. This is done by applying a 2D Gaussian function to the image with a given standard deviation $\sigma$. Although this type of blurring can improve the quality of an image, it can also alter the distribution of noise in the image. Therefore, when restoring an image, the blurring must be removed along with the distortion. In our training set, we only applied random $\sigma$ values between 0 and 1 pixel to the distorted images.
\textbf{Annotations} are added to an image to provide additional information or to highlight specific areas of the image. These can include text, shapes, and arrows, and may be added by the software or by the user. When creating training image pairs, we model the annotations by adding the same random annotations at the same pixel location in both the ground-truth and distorted images.
\textbf{Cropping} is a type of post-processing distortion that involves removing one or more areas of an image. This can be done manually by the user or automatically in a processing workflow, such as after the image has been shifted, rotated or aligned. The removed areas are usually filled in with a constant value or the median of the image's value range. When creating training image pairs, we model this process by randomly replacing the intensity value in a randomly selected area in both images. The selected area is typically outside a central square or rectangle, such as $50\%$ of the total image area, to mimic the fact that cropping is typically not applied to the central region, which may already be adjusted to show the main feature of interest.

\section*{CODE AND DATA AVAILABILITY}
All of the trained models, alongside example scripts for using them, are available on the github repository \url{https://github.com/Ivanlh20/r_em}. Additional material may be provided by the authors upon reasonable request.

\section*{ACKNOWLEDGEMENTS}
This work was supported by the European Research Council (Grant 770887 375 PICOMETRICS to S.V.A. and Grant 823717 ESTEEM3). The authors acknowledge financial support from the Research Foundation Flanders (FWO, Belgium) through project fundings (G.0346.21N and EOS 40007495). S.V.A. acknowledges funding from the University of Antwerp Research fund (BOF). The authors thank Lukas Grünewald for data acquisition and support for figure \ref{fig:concl_n2v}.

\section*{AUTHOR CONTRIBUTIONS}
I.L. and S.V.A. designed the study. I.L. created the mathematical models for the undistorted and distorted EM images, implemented, trained, and evaluated the NN models. T.F. conducted quantitative analysis of STEM images for the models. All authors contributed to the planning and execution of the research, discussed the results, and helped write the manuscript.

\section*{COMPETING INTERESTS}
The authors declare no competing interests.

\section*{ADDITIONAL INFORMATION}
 {\bf Supplementary information} is available for this article. \\

\noindent {\bf Correspondence} and requests for materials should be addressed to I.L.(Ivan.Lobato@uantwerpen.be) or S.V.A. (Sandra.Vanaert@uantwerpen.be).

\bibliographystyle{unsrt} 
\bibliography{references} 

\begin{thebibliography}{10}

\bibitem{NCDKEA04}
P.~D. Nellist, M.~F. Chisholm, N.~Dellby, O.~L. Krivanek, M.~F. Murfitt, Z.~S.
  Szilagyi, A.~R. Lupini, A.~Borisevich, W.~H.~Sides Jr., and S.~J. Pennycook.
\newblock {Direct Sub-Angstrom Imaging of a Crystal Lattice}.
\newblock {\em Science}, 305:1741, 2004.

\bibitem{Joy2005}
David~C Joy.
\newblock {The aberration corrected SEM}.
\newblock In {\em AIP Conference Proceedings}, volume 788, pages 535--542,
  2005.

\bibitem{KFBEA08}
C.~Kisielowski, B.~Freitag, M.~Bischoff, H.~van Lin, S.~Lazar, G.~Knippels,
  P.~Tiemeijer, M.~van~der Stam, S.~von Harrach, M.~Stekelenburg, M.~Haider,
  S.~Uhlemann, H.~M\"{u}ller, P.~Hartel, B.~Kabius, D.~Miller, I.~Petrov, E.~A.
  Olson, T.~Donchev, E.~A. Kenik, A.~R. Lupini, J.~Bentley, S.~J. Pennycook,
  I.~M. Anderson, A.~M. Minor, A.~K. Schmid, T.~Duden, V.~Radmilovic, Q.~M.
  Ramasse, M.~Watanabe, R.~Erni, E.~A. Stach, P.~Denes, and U.~Dahmen.
\newblock {Detection of Single Atoms and Buried Defects in Three Dimensions by
  Aberration-Corrected Electron Microscope with {0. 5\AA}-Information Limit}.
\newblock {\em Microscopy and Microanalysis}, 14:469--477, 2008.

\bibitem{MKTGFW06}
D.~A. Muller, E.~J. Kirkland, M.~G. Thomas, J.~L. Grazul, L.~Fitting, and
  M.~Weyland.
\newblock {Room design for high-performance electron microscopy}.
\newblock {\em Ultramicroscopy}, 106:1033--1040, 2006.

\bibitem{Jones2017}
Lewys Jones, Sigurd Wenner, Magnus Nord, Per~Harald Ninive, Ole~Martin
  L{\o}vvik, Randi Holmestad, and Peter~D. Nellist.
\newblock {Optimising multi-frame ADF-STEM for high-precision atomic-resolution
  strain mapping}.
\newblock {\em Ultramicroscopy}, 179:57--62, aug 2017.

\bibitem{Zuo2000}
J.~M. Zuo.
\newblock {Electron Detection Characteristics of a Slow-Scan CCD Camera,
  Imaging Plates and Film, and Electron Image Restoration}.
\newblock {\em Microscopy Research and Technique}, 49:245--268, 2000.

\bibitem{Faruqi2000}
A.~R. Faruqi and Sriram Subramaniam.
\newblock {CCD detectors in high-resolution biological electron microscopy},
  2000.

\bibitem{Ishikawa2014}
Ryo Ishikawa, Andrew~R. Lupini, Scott~D. Findlay, and Stephen~J. Pennycook.
\newblock {Quantitative annular dark field electron microscopy using single
  electron signals}.
\newblock {\em Microscopy and Microanalysis}, 20(1):99--110, 2014.

\bibitem{BBBDSV14}
B.~Berkels, P.~Binev, D.~A. Blom, W.~Dahmen, R.~C. Sharpley, and T.~Vogt.
\newblock Optimized imaging using non-rigid registration.
\newblock {\em Ultramicroscopy}, 138:46--56, 2014.

\bibitem{JYPMABCN15}
L.~Jones, H.~Yang, T.~J. Pennycook, M.~S.~J. Marshall, S.~Van Aert, N.~D.
  Browning, M.~R. Castell, and P.~D. Nellist.
\newblock Smart align-a new tool for robust non-rigid registration of scanning
  microscope data.
\newblock {\em Advanced Structural and Chemical Imaging}, 1:8, 2015.

\bibitem{LJPN13}
L.~Jones and P.~D. Nellist.
\newblock Identifying and correcting scan noise and drift in the scanning
  transmission electron microscope.
\newblock {\em Microscopy and Microanalysis}, 19:1050--1060, 2013.

\bibitem{BBLR12}
N.~Braidy, Y.~Le Bouar, S.~Lazar, and C.~Ricolleau.
\newblock Correcting scanning instabilities from images of periodic structures.
\newblock {\em Ultramicroscopy}, 118:67--76, 2012.

\bibitem{BRGBS10}
James~P. Buban, Quentin Ramasse, Bryant Gipson, Nigel~D. Browning, and Henning
  Stahlberg.
\newblock High-resolution low-dose scanning transmission electron microscopy.
\newblock {\em Journal of Electron Microscopy}, 59(2):103--112, 2010.

\bibitem{JCVD89}
A.~De Jong, W.~Coene, and D.~Van Dyck.
\newblock Image processing of hrtem images with non-periodic features.
\newblock {\em Ultramicroscopy}, 27(1):53--65, 1989.

\bibitem{LDM96}
L.~D. Marks.
\newblock Wiener-filter enhancement of noisy hrem images.
\newblock {\em Ultramicroscopy}, 62:43--52, 1996.

\bibitem{KTS12}
H.~S. Kushwaha, S.~Tanwar, K.~S. Rathore, and S.~Srivastava.
\newblock De-noising filters for tem (transmission electron microscopy) image
  of nanomaterials.
\newblock In {\em Proceedings of the 2012 Second International Con-ference on
  Advanced Computing \& Communication Technologies (ACCT)}, 2012.

\bibitem{HD15}
Hongchu Du.
\newblock A nonlinear filtering algorithm for denoising hr(s)tem micrographs.
\newblock {\em Ultramicroscopy}, 151:62--67, 2015.

\bibitem{Dabov2007}
Kostadin Dabov, Alessandro Foi, Vladimir Katkovnik, and Karen Egiazarian.
\newblock Image denoising by sparse 3-d transform-domain collaborative
  filtering.
\newblock {\em IEEE Transactions on Image Processing}, 16:2080--2095, 8 2007.

\bibitem{MBDVYB15}
Niklas Mevenkamp, Peter Binev, Wolfgang Dahmen, Paul~M. Voyles, Andrew~B.
  Yankovich, and Benjamin Berkels.
\newblock Poisson noise removal from high-resolution stem images based on
  periodic block matching.
\newblock {\em Advanced Structural and Chemical Imaging}, 1:1--19, 2015.

\bibitem{KAYNMI10}
Koji Kimoto, Toru Asaka, Xiuzhen Yu, Takuro Nagai, Yoshio Matsui, and Kazuo
  Ishizuka.
\newblock Local crystal structure analysis with several picometer precision
  using scanning transmission electron microscopy.
\newblock {\em Ultramicroscopy}, 110:778--782, 2010.

\bibitem{Huang2017}
Gao Huang, Zhuang Liu, Laurens {Van Der Maaten}, and Kilian~Q. Weinberger.
\newblock {Densely connected convolutional networks}.
\newblock In {\em Proceedings - 30th IEEE Conference on Computer Vision and
  Pattern Recognition, CVPR 2017}, volume 2017-Janua, pages 2261--2269.
  Institute of Electrical and Electronics Engineers Inc., aug 2017.

\bibitem{Chen2018}
Liang~Chieh Chen, Yukun Zhu, George Papandreou, Florian Schroff, and Hartwig
  Adam.
\newblock {Encoder-decoder with atrous separable convolution for semantic image
  segmentation}.
\newblock In {\em Lecture Notes in Computer Science (including subseries
  Lecture Notes in Artificial Intelligence and Lecture Notes in
  Bioinformatics)}, volume 11211 LNCS, pages 833--851. Springer Verlag, feb
  2018.

\bibitem{Kim2019}
Dong~Wook Kim, Jae~Ryun Chung, and Seung~Won Jung.
\newblock {GRDN:Grouped residual dense network for real image denoising and
  GAN-based real-world noise modeling}.
\newblock In {\em IEEE Computer Society Conference on Computer Vision and
  Pattern Recognition Workshops}, volume 2019-June, pages 2086--2094, 2019.

\bibitem{Cheng2018}
Ziang Cheng, Shaodi You, Viorela Ila, and Hongdong Li.
\newblock Semantic {{Single-Image Dehazing}}, April 2018.

\bibitem{Luo}
Zhijian Luo, Siyu Chen, and Yuntao Qian.
\newblock A {{Deep Optimization Approach}} for {{Image Deconvolution}}, April
  2019.

\bibitem{Wang2019}
Xintao Wang, Ke~Yu, Shixiang Wu, Jinjin Gu, Yihao Liu, Chao Dong, Yu~Qiao, and
  Chen~Change Loy.
\newblock {ESRGAN: Enhanced super-resolution generative adversarial networks}.
\newblock In {\em Lecture Notes in Computer Science (including subseries
  Lecture Notes in Artificial Intelligence and Lecture Notes in
  Bioinformatics)}, volume 11133 LNCS, pages 63--79. Springer Verlag, sep 2019.

\bibitem{VNEH10}
Vinod Nair and Geoffrey~E. Hinton.
\newblock Rectified linear units improve restricted boltzmann machines.
\newblock In {\em International Conference on Machine Learning}, pages
  807--814, 2010.

\bibitem{SHKSS14}
Nitish Srivastava, Geoffrey Hinton, Alex Krizhevsky, Ilya Sutskever, and Ruslan
  Salakhutdinov.
\newblock Dropout: A simple way to prevent neural networks from overfitting.
\newblock {\em Journal of Machine Learning Research}, 15:1929--1958, 2014.

\bibitem{Ioffe2015}
Sergey Ioffe and Christian Szegedy.
\newblock {Batch normalization: Accelerating deep network training by reducing
  internal covariate shift}.
\newblock In {\em 32nd International Conference on Machine Learning, ICML
  2015}, volume~1, pages 448--456. International Machine Learning Society
  (IMLS), feb 2015.

\bibitem{Wang2020a}
Feng Wang, Alberto Eljarrat, Johannes Müller, Trond~R. Henninen, Rolf Erni,
  and Christoph~T. Koch.
\newblock Multi-resolution convolutional neural networks for inverse problems.
\newblock {\em Scientific Reports}, 10:1--11, 12 2020.

\bibitem{Wang2020b}
Feng Wang, Trond~R Henninen, Debora Keller, and Rolf Erni.
\newblock Noise2atom: unsupervised denoising for scanning transmission electron
  microscopy images.
\newblock {\em Applied Microscopy}, 50, 2020.

\bibitem{Mohan2022}
Sreyas Mohan, Ramon Manzorro, Joshua~L. Vincent, Binh Tang, Dev~Y. Sheth,
  Eero~P. Simoncelli, David~S. Matteson, Peter~A. Crozier, and Carlos
  {Fernandez-Granda}.
\newblock Deep {{Denoising}} for {{Scientific Discovery}}: {{A Case Study}} in
  {{Electron Microscopy}}.
\newblock {\em IEEE Transactions on Computational Imaging}, 8:585--597, 2022.

\bibitem{Lin2021}
Ruoqian Lin, Rui Zhang, Chunyang Wang, Xiao-Qing Yang, and Huolin~L. Xin.
\newblock {{TEMImageNet}} training library and {{AtomSegNet}} deep-learning
  models for high-precision atom segmentation, localization, denoising, and
  deblurring of atomic-resolution images.
\newblock {\em Scientific Reports}, 11(1):5386, March 2021.

\bibitem{Kaiser2023}
Eurika Kaiser, Nathan Kutz, Steven~L Brunton, Laura Gambini, Tiarnan Mullarkey,
  Lewys Jones, and Stefano Sanvito.
\newblock Machine-learning approach for quantified resolvability enhancement of
  low-dose stem data.
\newblock {\em Machine Learning: Science and Technology}, 4:015025, 2 2023.

\bibitem{Mirza2014}
Mehdi Mirza and Simon Osindero.
\newblock Conditional {{Generative Adversarial Nets}}, November 2014.

\bibitem{Ronneberger2015}
Olaf Ronneberger, Philipp Fischer, and Thomas Brox.
\newblock U-net: Convolutional networks for biomedical image segmentation.
\newblock {\em arXiv}, pages 1--8, 2015.

\bibitem{VanAert2009}
S.~{Van Aert}, J.~Verbeeck, R.~Erni, S.~Bals, M.~Luysberg, D.~Van Dyck, and
  G.~Van Tendeloo.
\newblock {Quantitative atomic resolution mapping using high-angle annular dark
  field scanning transmission electron microscopy}.
\newblock {\em Ultramicroscopy}, 109(10):1236--1244, sep 2009.

\bibitem{Martinez2014}
G~T Martinez, A.~{De Backer}, A~Rosenauer, J~Verbeeck, and S.~{Van Aert}.
\newblock {The effect of probe inaccuracies on the quantitative model-based
  analysis of high angle annular dark field scanning transmission electron
  microscopy images}.
\newblock {\em Micron}, 63(2014):57--63, 2014.

\bibitem{DeBacker2016}
A.~{De Backer}, K.~H.W. van~den Bos, W.~{Van den Broek}, J.~Sijbers, and
  S.~{Van Aert}.
\newblock {StatSTEM: An efficient approach for accurate and precise model-based
  quantification of atomic resolution electron microscopy images}.
\newblock {\em Ultramicroscopy}, 171(2016):104--116, 2016.

\bibitem{Krivanek2010a}
Ondrej~L. Krivanek, Matthew~F. Chisholm, Valeria Nicolosi, Timothy~J.
  Pennycook, George~J. Corbin, Niklas Dellby, Matthew~F. Murfitt,
  Christopher~S. Own, Zoltan~S. Szilagyi, Mark~P. Oxley, Sokrates~T.
  Pantelides, and Stephen~J. Pennycook.
\newblock Atom-by-atom structural and chemical analysis by annular dark-field
  electron microscopy.
\newblock {\em Nature 2010 464:7288}, 464(7288):571--574, March 2010.

\bibitem{Yamashita2018a}
Shunsuke Yamashita, Jun Kikkawa, Keiichi Yanagisawa, Takuro Nagai, Kazuo
  Ishizuka, and Koji Kimoto.
\newblock Atomic number dependence of {{Z}} contrast in scanning transmission
  electron microscopy.
\newblock {\em Scientific Reports}, 8(1):1--7, 2018.

\bibitem{Lobato2015}
I.~Lobato and D.~{Van Dyck}.
\newblock {MULTEM: A new multislice program to perform accurate and fast
  electron diffraction and imaging simulations using Graphics Processing Units
  with CUDA}.
\newblock {\em Ultramicroscopy}, 156:9--17, sep 2015.

\bibitem{Lobato2016}
I.~Lobato, S.~van Aert, and J.~Verbeeck.
\newblock {Progress and new advances in simulating electron microscopy datasets
  using MULTEM}.
\newblock {\em Ultramicroscopy}, 168:17--27, sep 2016.

\bibitem{Hull2011}
Derek Hull and D.~J. Bacon.
\newblock {\em Introduction to Dislocations}.
\newblock Butterworth-Heinemann, 2011.

\bibitem{Hirel2015}
Pierre Hirel.
\newblock {Atomsk: A tool for manipulating and converting atomic data files}.
\newblock {\em Computer Physics Communications}, 197:212--219, dec 2015.

\bibitem{VanAert2013}
S.~{Van Aert}, A.~{De Backer}, G.~T. Martinez, B.~Goris, S.~Bals, G.~{Van
  Tendeloo}, and A.~Rosenauer.
\newblock {Procedure to count atoms with trustworthy single-atom sensitivity}.
\newblock {\em Physical Review B - Condensed Matter and Materials Physics},
  87(6), feb 2013.

\bibitem{ArslanIrmak2021}
Ece Arslan~Irmak, Pei Liu, Sara Bals, and Sandra Van~Aert.
\newblock {{3D Atomic Structure}} of {{Supported Metallic Nanoparticles
  Estimated}} from {{2D ADF STEM Images}}: {{A Combination}} of
  {{Atom-Counting}} and a {{Local Minima Search Algorithm}}.
\newblock {\em Small Methods}, 5(12):2101150, 2021.

\bibitem{DeBacker2022}
Annick De~Backer, Sandra Van~Aert, Christel Faes, Ece Arslan~Irmak, Peter~D.
  Nellist, and Lewys Jones.
\newblock Experimental reconstructions of {{3D}} atomic structures from
  electron microscopy images using a {{Bayesian}} genetic algorithm.
\newblock {\em npj Computational Materials}, 8(1):1--8, October 2022.

\bibitem{Ede2020}
Jeffrey~M. Ede.
\newblock Warwick electron microscopy datasets.
\newblock {\em Machine Learning: Science and Technology}, 1:045003, 9 2020.

\bibitem{Aversa2018}
Rossella Aversa, Mohammad~Hadi Modarres, Stefano Cozzini, Regina Ciancio, and
  Alberto Chiusole.
\newblock The first annotated set of scanning electron microscopy images for
  nanoscience.
\newblock {\em Scientific Data}, 5:1--10, 8 2018.

\bibitem{Altantzis2019}
Thomas Altantzis, Ivan Lobato, Annick {De Backer}, Armand B{\'{e}}ch{\'{e}},
  Yang Zhang, Shibabrata Basak, Mauro Porcu, Qiang Xu, Ana
  S{\'{a}}nchez-Iglesias, Luis~M. Liz-Marz{\'{a}}n, Gustaaf {Van Tendeloo},
  Sandra {Van Aert}, and Sara Bals.
\newblock {Three-Dimensional Quantification of the Facet Evolution of Pt
  Nanoparticles in a Variable Gaseous Environment}.
\newblock {\em Nano Letters}, 19(1):477--481, jan 2019.

\bibitem{Amini2018}
Mozhgan~N. Amini, Thomas Altantzis, Ivan Lobato, Marek Grzelczak, Ana
  {S{\'a}nchez-Iglesias}, Sandra Van~Aert, Luis~M. {Liz-Marz{\'a}n}, Bart
  Partoens, Sara Bals, and Erik~C. Neyts.
\newblock Understanding the {{Effect}} of {{Iodide Ions}} on the {{Morphology}}
  of {{Gold Nanorods}}.
\newblock {\em Particle \& Particle Systems Characterization}, 35(8):1800051,
  2018.

\bibitem{Krull2019}
Alexander Krull, Tim-Oliver Buchholz, and Florian Jug.
\newblock {{Noise2Void}} - {{Learning Denoising From Single Noisy Images}}.
\newblock In {\em Proceedings of the {{IEEE}}/{{CVF Conference}} on {{Computer
  Vision}} and {{Pattern Recognition}}}, pages 2129--2137, 2019.

\bibitem{Cayado2022}
Pablo Cayado, Lukas Gr{\"u}newald, Manuela Erbe, Jens H{\"a}nisch, Dagmar
  Gerthsen, and Bernhard Holzapfel.
\newblock Critical current density improvement in {{CSD-grown}} high-entropy
  {{REBa}}
  \textsubscript{2}{{Cu}}\textsubscript{3}{{O}}\textsubscript{7-{$\delta$}}
  films.
\newblock {\em RSC Advances}, 12(44):28831--28842, 2022.

\bibitem{Gruenewald2022}
Lukas Gr{\"u}newald.
\newblock {\em {Electron Microscopic Investigation of Superconducting Fe- and
  Cu-based Thin Films}}.
\newblock PhD thesis, Laboratorium f\"ur Elektronenmikroskopie (LEM),
  Karlsruher Institut f\"ur Technologie (KIT), 2022.

\bibitem{Molina-Luna2015}
Leopoldo {Molina-Luna}, Michael Duerrschnabel, Stuart Turner, Manuela Erbe,
  Gerardo~T Martinez, Sandra Van~Aert, Bernhard Holzapfel, and Gustaaf
  Van~Tendeloo.
\newblock Atomic and electronic structures of
  {{BaHfO}}{\textsubscript{3}}-doped {{TFA-MOD-derived YBa}}
  {\textsubscript{2}}{{Cu}}{\textsubscript{3}}{{O}}{\textsubscript{7-
  {$\delta$}}} thin films.
\newblock {\em Superconductor Science and Technology}, 28(11):115009, November
  2015.

\bibitem{Zhao2017}
Hang Zhao, Orazio Gallo, Iuri Frosio, and Jan Kautz.
\newblock {Loss Functions for Image Restoration With Neural Networks}.
\newblock {\em IEEE Transactions on Computational Imaging}, 3(1):47--57, 2016.

\bibitem{Isola2016}
Phillip Isola, Jun-Yan Zhu, Tinghui Zhou, and Alexei~A. Efros.
\newblock {Image-to-Image Translation with Conditional Adversarial Networks}.
\newblock {\em Proceedings - 30th IEEE Conference on Computer Vision and
  Pattern Recognition, CVPR 2017}, 2017-Janua:5967--5976, nov 2016.

\bibitem{Goodfellow}
Ian~J. Goodfellow, Jean {Pouget-Abadie}, Mehdi Mirza, Bing Xu, David
  {Warde-Farley}, Sherjil Ozair, Aaron Courville, and Yoshua Bengio.
\newblock Generative {{Adversarial Networks}}, June 2014.

\bibitem{Zhang2018_1}
Yulun Zhang, Yapeng Tian, Yu~Kong, Bineng Zhong, and Yun Fu.
\newblock {Residual Dense Network for Image Restoration}.
\newblock In {\em Proceedings of the IEEE Computer Society Conference on
  Computer Vision and Pattern Recognition}, pages 2472--2481, dec 2018.

\bibitem{Ignatov2019}
Andrey Ignatov, Radu Timofte, Xiaochao Qu, Xingguang Zhou, Ting Liu, Pengfei
  Wan, Syed~Waqas Zamir, Aditya Arora, Salman Khan, Fahad~Shahbaz Khan, Ling
  Shao, Dongwon Park, Se~Young Chun, Pablo~Navarrete Michelini, Hanwen Liu, Dan
  Zhu, Zhiwei Zhong, Xianming Liu, Junjun Jiang, Debin Zhao, Muhammad Haris,
  Kazutoshi Akita, Tomoki Yoshida, Greg Shakhnarovich, Norimichi Ukita, Jie
  Liu, Cheolkon Jung, Raimondo Schettini, Simone Bianco, Claudio Cusano, Flavio
  Piccoli, Pengju Liu, Kai Zhang, Jingdong Liu, Jiye Liu, Hongzhi Zhang,
  Wangmeng Zuo, Nelson Chong~Ngee Bow, Lai~Kuan Wong, John See, Jinghui Qin,
  Lishan Huang, Yukai Shi, Pengxu Wei, Wushao Wen, Liang Lin, Zheng Hui, Xiumei
  Wang, Xinbo Gao, Kanti Kumari, Vikas~Kumar Anand, Mahendra Khened, and
  Ganapathy Krishnamurthi.
\newblock {NTIRE 2019 challenge on image enhancement: Methods and results}.
\newblock In {\em IEEE Computer Society Conference on Computer Vision and
  Pattern Recognition Workshops}, volume 2019-June, pages 2224--2232, 2019.

\bibitem{Woo2018}
Sanghyun Woo, Jongchan Park, Joon-Young Lee, and In~So Kweon.
\newblock {CBAM: Convolutional Block Attention Module}.
\newblock {\em Lecture Notes in Computer Science (including subseries Lecture
  Notes in Artificial Intelligence and Lecture Notes in Bioinformatics)}, 11211
  LNCS:3--19, jul 2018.

\bibitem{Plotz2018}
Tobias Pl{\"{o}}tz and Stefan Roth.
\newblock {Neural Nearest Neighbors Networks}.
\newblock {\em Advances in Neural Information Processing Systems},
  2018-December:1087--1098, oct 2018.

\bibitem{Zhang2017}
Kai Zhang, Wangmeng Zuo, Yunjin Chen, Deyu Meng, and Lei Zhang.
\newblock {Beyond a Gaussian denoiser: Residual learning of deep CNN for image
  denoising}.
\newblock {\em IEEE Transactions on Image Processing}, 26(7):3142--3155, jul
  2017.

\bibitem{Li2015}
Zuhe Li, Yangyu Fan, and Weihua Liu.
\newblock {The effect of whitening transformation on pooling operations in
  convolutional autoencoders}.
\newblock {\em Eurasip Journal on Advances in Signal Processing},
  2015(1):1--11, dec 2015.

\bibitem{Jolicoeur-Martineau2019}
Alexia Jolicoeur-Martineau.
\newblock {The relativistic discriminator: A key element missing from standard
  GaN}.
\newblock In {\em 7th International Conference on Learning Representations,
  ICLR 2019}. International Conference on Learning Representations, ICLR, jul
  2019.

\bibitem{Jarrett}
Kevin Jarrett, Koray Kavukcuoglu, Marc'Aurelio Ranzato, and Yann LeCun.
\newblock What is the best multi-stage architecture for object recognition?
\newblock In {\em 2009 {{IEEE}} 12th {{International Conference}} on {{Computer
  Vision}}}, pages 2146--2153, September 2009.

\bibitem{VanAert2011}
Sandra {Van Aert}, Kees~J. Batenburg, Marta~D. Rossell, Rolf Erni, and Gustaaf
  {Van Tendeloo}.
\newblock {Three-dimensional atomic imaging of crystalline nanoparticles}.
\newblock {\em Nature}, 470(7334):374--377, feb 2011.

\bibitem{Beyer2016}
A~Beyer, R~Straubinger, J~Belz, and K~Volz.
\newblock {Local sample thickness determination via scanning transmission
  electron microscopy defocus series}.
\newblock {\em Journal of Microscopy}, 262(2):171--177, 2016.

\bibitem{Lobato2014}
I.~Lobato and D.~{Van Dyck}.
\newblock {An accurate parameterization for scattering factors, electron
  densities and electrostatic potentials for neutral atoms that obey all
  physical constraints}.
\newblock {\em Acta Crystallographica Section A: Foundations and Advances},
  70:636--649, nov 2014.

\bibitem{Hartel1996}
P.~Hartel, H.~Rose, and C.~Dinges.
\newblock {Conditions and reasons for incoherent imaging in STEM}.
\newblock {\em Ultramicroscopy}, 63(2):93--114, jun 1996.

\bibitem{Chew1985}
L.~Paul Chew and Robert~L. Drysdale.
\newblock Voronoi diagrams based or convex distance functions.
\newblock {\em Proceedings of the 1st Annual Symposium on Computational
  Geometry, SCG 1985}, pages 235--244, 6 1985.

\bibitem{Spence1988}
J.~C.~H. Spence and J.~M. Zuo.
\newblock {Large dynamic range, parallel detection system for electron
  diffraction and imaging}.
\newblock {\em Review of Scientific Instruments}, 59(9):2102--2105, sep 1988.

\bibitem{Tietz2008}
HR~Tietz.
\newblock Design and characterization of 64 megapixel fiber optic coupled cmos
  detector for transmission electron microscopy.
\newblock {\em Microscopy and Microanalysis}, 14(S2):804–805, 2008.

\bibitem{Clough2014}
R.~N. Clough, G.~Moldovan, and A.~I. Kirkland.
\newblock Direct {{Detectors}} for {{Electron Microscopy}}.
\newblock {\em Journal of Physics: Conference Series}, 522(1):012046, June
  2014.

\bibitem{Thust2009}
A.~Thust.
\newblock {High-resolution transmission electron microscopy on an absolute
  contrast scale}.
\newblock {\em Physical Review Letters}, 102(22), jun 2009.

\bibitem{Vulovic2010}
M.~Vulovic, B.~Rieger, L.~J. {Van Vliet}, A.~J. Koster, and R.~B.~G. Ravelli.
\newblock {A toolkit for the characterization of CCD cameras for transmission
  electron microscopy}.
\newblock {\em Acta Crystallographica Section D: Biological Crystallography},
  66(1):97--109, dec 2010.

\bibitem{Kodak2005}
Kodak.
\newblock {CCD Image Sensor Noise Sources}.
\newblock {\em Application Note}, 2005.

\bibitem{Konnik2014}
Mikhail Konnik and James Welsh.
\newblock High-level numerical simulations of noise in {{CCD}} and {{CMOS}}
  photosensors: Review and tutorial, December 2014.

\bibitem{Gow2007}
Ryan~D. Gow, David Renshaw, Keith Findlater, Lindsay Grant, Stuart~J. McLeod,
  John Hart, and Robert~L. Nicol.
\newblock {A comprehensive tool for modeling CMOS image-sensor-noise
  performance}.
\newblock {\em IEEE Transactions on Electron Devices}, 54(6):1321--1329, 2007.

\bibitem{Irie2008}
K.~Irie, A.~E. McKinnon, K.~Unsworth, and I.~M. Woodhead.
\newblock A model for measurement of noise in ccd digital-video cameras.
\newblock {\em Measurement Science and Technology}, 19:045207, 3 2008.

\bibitem{ZINGER1961}
A.~ZINGER.
\newblock Detection of best and outlying normal populations with known
  variances.
\newblock {\em Biometrika}, 48:457--461, 12 1961.

\bibitem{Yule1927}
G.~Udny Yule.
\newblock {On a Method of Investigating Periodicities in Disturbed Series}.
\newblock {\em Philosophical Transactions of the Royal Society of London},
  226(Series A):167--298, 1927.

\bibitem{GW31}
G.~Walker.
\newblock On periodicity in series of related terms.
\newblock {\em Proceedings of the Royal Society A}, 131:518--32, 1931.

\bibitem{Burle1984}
Burle Industries.
\newblock {\em Photomultiplier Handbook}.
\newblock Burle Technologies Inc., 1980.

\bibitem{Mittelberger2018}
Andreas Mittelberger, Christian Kramberger, and Jannik~C. Meyer.
\newblock Software electron counting for low-dose scanning transmission
  electron microscopy.
\newblock {\em Ultramicroscopy}, 188:1--7, 2018.

\bibitem{Mullarkey2021}
Tiarnan Mullarkey, Clive Downing, and Lewys Jones.
\newblock Development of a {{Practicable Digital Pulse Read-Out}} for
  {{Dark-Field STEM}}.
\newblock {\em Microscopy and Microanalysis}, 27(1):99--108, February 2021.

\bibitem{KrivanekCD2008}
O.~L. Krivanek, G.~J. Corbin, N.~Dellby, B.~F. Elston, R.~J. Keyse, M.~F.
  Murfitt, C.~S. Own, Z.~S. Szilagyi, and J.~W. Woodruff.
\newblock An electron microscope for the aberration-corrected era.
\newblock {\em Ultramicroscopy}, 108:179--195, 2 2008.

\bibitem{PKJB14}
Diederik~P. Kingma and Jimmy Ba.
\newblock Adam: A method for stochastic optimization.
\newblock {\em arXiv:1412. 6980}, 2014.

\bibitem{Vaswani2017}
Ashish Vaswani, Noam Shazeer, Niki Parmar, Jakob Uszkoreit, Llion Jones,
  Aidan~N. Gomez, {\L}ukasz Kaiser, and Illia Polosukhin.
\newblock {Attention is all you need}.
\newblock In {\em Advances in Neural Information Processing Systems}, volume
  2017-Decem, pages 5999--6009, jun 2017.

\bibitem{Popel2018}
Martin Popel and Ondřej Bojar.
\newblock {Training Tips for the Transformer Model}.
\newblock {\em The Prague Bulletin of Mathematical Linguistics}, 110(1):43--70,
  mar 2018.

\end{thebibliography}
\clearpage
\include{supplementary.tex}
\end{document}